\documentclass[ aps, twocolumn,secnumarabic, superscriptaddress,
nofootinbib,
 amsmath,amssymb,
floatfix,
]{revtex4-1}
\usepackage{ulem}
\usepackage{physics}
\usepackage{epstopdf}
\usepackage{color}
\usepackage{hyperref}
\usepackage{bm}
\usepackage{graphicx}% Include figure files
\usepackage{dcolumn}% Align table columns on decimal point
\usepackage{bm}% bold math
\usepackage{float}
\renewcommand{\vec}[1]{\mathbf{#1}}

\newcommand{\no}{\noindent}

\renewcommand{\k}{\mathbf{k}}
\renewcommand{\r}{\mathbf{r}}

\newcommand{\vv}{\boldsymbol}

\def\res{\mathop{Res}}

\newcommand{\NH}{\hat{\text H}}
\newcommand{\HH}{\hat{\mathbb H}}

\newcommand{\G}{{\text G}}
\newcommand{\GG}{{\mathbb G}}

\newcommand{\PT}{$\mathcal{P} \mathcal{T}$}

\newcommand*{\figuretitle}[1]{%
    {\centering%   <--------  will only affect the title because of the grouping (by the
    \textbf{#1}%              braces before \centering and behind \medskip). If you remove
    \par\medskip}%            these braces the whole body of a {figure} env will be centered.
}

\begin{document}

\title{Non-Hermitian Boundary Modes}

 \author{Dan S. Borgnia}
 \email[]{dborgnia@g.harvard.edu}
 \affiliation{Department of Physics, Harvard University, Cambridge, MA 02138} 
 \author{Alex Jura Kruchkov}
 \affiliation{Department of Physics, Harvard University, Cambridge, MA 02138} 
 \author{Robert-Jan Slager} 
 \email[]{rjslager@g.harvard.edu}
 \affiliation{Department of Physics, Harvard University, Cambridge, MA 02138}

\date{\today}

\begin{abstract}
We consider conditions for the existence of boundary modes in non-Hermitian systems with edges of arbitrary co-dimension. Through a universal formulation of formation criteria for boundary modes in terms of local Green functions, we outline a generic perspective on the appearance of such modes and generate corresponding dispersion relations. In the process, we explain the skin effect in both topological and non-topological systems, exhaustively generalizing bulk-boundary correspondence in the presence of non-Hermiticity. This is accomplished via a doubled Green's function, inspired by doubled Hamiltonian methods used to classify Floquet and, more recently, non-Hermitian topological phases. Our work constitutes a general tool, as well as, a unifying perspective for this rapidly evolving field. Indeed, as a concrete application we find that our method can expose novel non-Hermitian topological regimes beyond the reach of previous methods. 
\end{abstract}

\maketitle

{\bf 
The relaxation of Hermiticity allows new symmetries and promises novel topological phases \cite{2018arXiv181210490Z,Kawabata2018,kawabata2019topological,kawabata2018symmetry,leykam2017edge, PhysRevLett.120.146402,Yao2018,PhysRevB.84.205128,lee2018tidal}. Recent experiments  \cite{Ganainy2018,Guo2009,Ruter2010,Zeuner2015, Zhoueaap9859} observe generalizations of concepts from Hermitian systems \cite{Shen2018,Gong2018}, but depend on fine-tuned gain and loss symmetries, e.g. \PT symmetry \cite{Bender1998,doi:10.1063/1.1418246,PhysRevE.59.6433}. Some aspects a priori appear unphysical, such as the pile-up of bulk states at system edges, the \textit{non-Hermitian skin effect} \cite{PhysRevLett.121.136802,alvarez2018topological,Yao2018,xiong2018does,lee2018anatomy}, which suggests an absence of \textit{bulk-boundary correspondences} \cite{PhysRevLett.121.136802,PhysRevLett.120.146402,Kunst2018a,Kunst2018b,2019arXiv190100010H,jin2018bulk}-- boundary modes reflecting topological degeneracy in the bulk spectrum \cite{Clas1a, Clas1b, Clas1c, Clas2, PhysRevB.90.165114, Clas3, PhysRevB.98.024310, Clas4, Clas5} that lay the foundation for the classification and observation of topological phases \cite{Slager2015, PhysRevLett.71.3697,Rhim2018,Clas1b, PhysRevB.93.245406}.
Here, we introduce a general framework, resting on universal Green's function classification technology \cite{Slager2015}, to formulate an exhaustive description of possible boundary modes and their connection to bulk topology. Through this physical, experimentally accessible, observable, we generalize notions of bulk boundary correspondence, explain the non-Hermitian skin effect, and uncover novel intermediate regimes. An otherwise Hermitian system is tuned from one non-trivial topological phase to another by adding generic non-Hermiticity, suggesting novel properties of real systems under open conditions.}\\
%%%%%%%%%%%%%%%%%%%%%%%%%%%%%%%%%%%%%%%%%%
\begin{figure*}
\figuretitle{Non-Hermitian Gap Conditions}
\includegraphics[width =\textwidth]{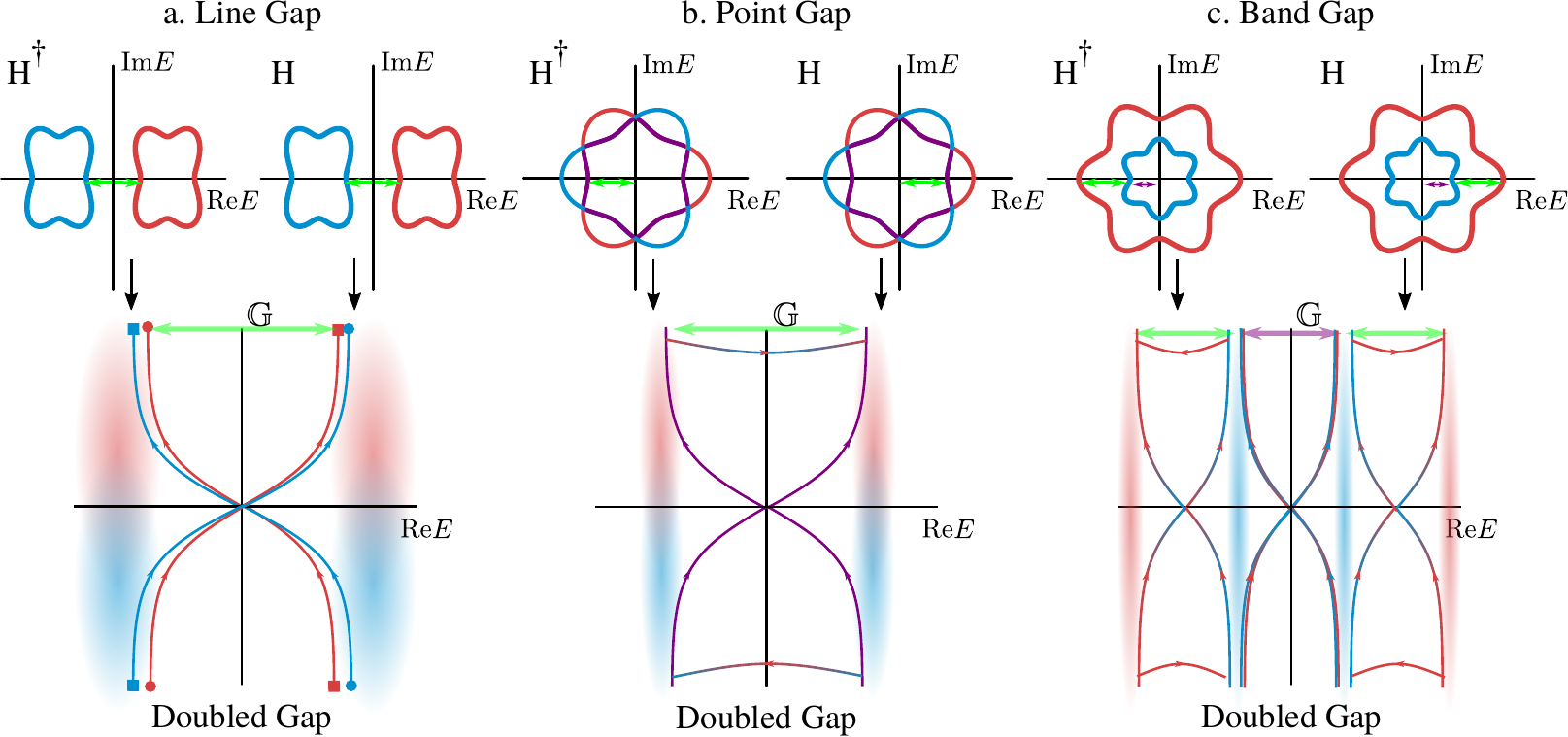}
\caption{Possible boundary modes with respect to the different band gap in non-Hermitian system. The top row presents Bloch bands on the complex plane generated by non-Hermitian Hamiltonians and their respective complex conjugates, showcasing band gaps (a,c) and a point gap (b). The second row illustrates the mapping of two (complex) single band gaps to a one (real) doubled band gap (Green/Purple arrows indicate gaps), $\NH^{\dagger} \rightarrow -E$ and $\NH \rightarrow +E$, with the blue-red smear representing the new degenerate bands. Curves crossing the gap depict a topological example of in-gap doubled Green's function $\GG$ eigenvalues, see Eq. \eqref{doubledgreen}. Gapless bands (b) become indistinguishable \cite{Shen2018}, and a line gap (a) protects zeros of $\GG$ crossing from red (blue) to blue (red) bands, while bands in (c) may have both point gap invariants (crossing zero in purple gap) and band gap invariants (red to blue) \cite{2018arXiv181210490Z,kawabata2018symmetry}.}
\label{figure1}
\end{figure*}
%%%%%%%%%%%%%%%%%%%%%%%%%%%%%%%%%%%%%%%%%%% 

\indent \textit{Non-Hermitian Band Topology} - 
Before addressing possible bulk boundary correspondences, we first briefly describe possible band gaps and topologies with respect to which these can be defined. Hermitian symmetry protected topological phases (SPTs) are classified with respect to a band gap \cite{Clas1a, Clas1b, Clas1c, Clas2, PhysRevB.90.165114, Clas3, PhysRevB.98.024310, Clas4, Clas5}. Non-Hermitian spectra are, however, generically complex. In the complex plane, a band gap can be either a line gap between two bands, Fig.~\ref{figure1}a, or the closed region between two bands, Fig.~\ref{figure1}c \cite{PhysRevLett.120.146402,Shen2018}. We define topology with respect to the gap center as is done for Hermitian SPTs. The presence of a protected line/band gap (Fig.~\ref{figure1}a,c) generates two disjoint regions of the complex plane just as a band gap does on the real axis. Thus, results from Hermitian band topology must hold for line/band gaps. 

\indent Line/band gaps are not, however, a complete picture of non-Hermitian band topology. Band topology may also be defined with respect to the band center, such as the inseparable bands \cite{Shen2018} in  Fig.~\ref{figure1}b \cite{2018arXiv181210490Z,kawabata2018symmetry,kawabata2019topological}. Point gaps are unique to non-Hermitian systems and may ascribe a net non-trivial system topology \cite{nonherm_green}. They generate a simply-connected punctured plane, whose topology is different from the two disjoint regions generated by a band gap. This geometric interpretation of different band gaps will guide the development of our formalism below. In fact, it implies the bulk topological invariant must change as a line gap closes into a point gap through the touching of isolated bands (Fig.~\ref{figure1}b)  \cite{kawabata2018symmetry,2018arXiv181210490Z,kawabata2019topological}. 

\indent Point gaps are generically present in addition to band/line gaps, as bands form closed loops on the complex plane. And, line to point gap closings are not fine tuned, only requiring bands enclose a single point, see Fig.~\ref{figure1}. Under open boundary conditions, however, bands are no longer guaranteed to close and point gaps are not generically preserved. Therefore, one should not expect traditional bulk boundary correspondence for point gap topological phases, as we demonstrate below.\\
%%%%%%%%%%%%%%%%%%%%%%%%%%%%%%%%%%%%%%%%%%%%%%%%%%%%%%%%%%%%%%%%%

\indent \textit{Diagnosing Edge-Localized Bound States -} All formation criteria for edge modes can be obtained from the in-gap zeros of Green's function restricted to the relevant edge of the system \cite{Slager2015,essin2011bulk,Rhim2018}. The idea is that the full Green's function of a system with an on-site potential $\mathcal{V}$ is given by \begin{eqnarray}\label{GFformalism}
\G(\omega,\vec{k}) = (1-\G_{0}(\omega,\vec{k})\mathcal{V})^{-1}\G_{0}(\omega,\vec{k}),
\end{eqnarray}
where we have suppressed indices on the possible matrix structure of $\mathcal{V}$ that is assumed to have no spatial dependence. Restricting the system to an edge, poles of the Green's function correspond to  \begin{equation}
	\det \left[ \G_0 (\omega,\k_\parallel,\r_\perp=0) \mathcal{V} - \mathbf{1} \right] = 0,
	\label{EvEquation}
\end{equation}
where $\k_{\parallel}$ and $\r_{\perp}$ refer to the unbroken in-plane momenta and perpendicular coordinates along an edge of arbitrary co-dimension (we focus on co-dimension-1) \cite{Slager2015}.

\indent Unlike their trivial counterparts, Hermitian topological phases must have in-gap solutions of Eq.~\ref{EvEquation} for any impurity strength approaching an edge, $|\mathcal{V}|\rightarrow\infty$  \cite{Slager2015}. Therefore, zeros of the restricted in-gap Green's function, $\G_0 (\epsilon,\k_\parallel,\r_\perp=0)$, correspond to topological edge-localized bound states. And, the locations of the Green's function zeros $\omega_*(\vec k_{\parallel})$ as a function of $\textbf{k}_{\parallel}$ form, by definition, a dispersion relation for the edge mode. Furthermore, counting Green's function zeros between bands determines the change in topological invariant from one band to the next \cite{bernevig2013topological}, see Supplemental Material \ref{GFchernappend} for more detail.\\
%%%%%%%%%%%%%%%%%%%%%%%%%%%%%%%%%%%%%%%%%%%%%%%%

\indent \textit{Doubled Green's Function }- We cannot directly apply the above formalism to a non-Hermitian system. Indeed, projection to an edge can destroy the point gap (Fig.~\ref{figure1} b). In close analogy to Floquet SPTs \cite{PhysRevB.96.155118}, given a non-Hermitian Hamiltonian $\text{H}$, we, therefore, define a Hermitian \textit{doubled Hamiltonian},
\begin{align}
\label{doubled_hamiltonian}
\HH 
=	\begin{pmatrix}
0 & \NH\\ \NH^{\dagger} & 0	
\end{pmatrix}. 
\end{align}
\noindent It maps the bands of $\NH$ to positive energies and the bands of $\NH^{\dagger}$ to negative energies, see Fig.~\ref{figure1}. Accordingly, the doubled topology encodes the sub-block, $\NH$, topology with respect to the gap at $E = 0$\cite{kawabata2018symmetry,2018arXiv181210490Z}. \\
\indent We define a \textit{doubled Green's function} corresponding to the doubled Hamiltonian,
\begin{align}
\label{doubledgreen}
\GG (\omega) =	
\frac{ \GG_{0}(\omega}
{1 - \hat{\mathbb{V}}\, \GG_{0}(\omega)}, \ \ \text{with} \ \ 
 \GG_{0}(\omega) =	
\frac{1}{\omega - \HH_0},
\end{align}
\noindent where $\hat{\mathbb{V}}$ is the doubled impurity potential, $\mathcal{V}(\vv r_{\perp}=0)$, and $\HH = \HH_{0} +\Hat{\mathbb{V}}$.  Being Hermitian, the topological edge states correspond to the zeros of the projected doubled Green's function, as before (Supplemental Material \ref{GFcorrappendix}). \\
\indent Next, we extract the physical meaning of these edge modes by relating doubled Green's function zeros to those of the single Green's function. Specifically, we parameterize the single Green's functions by $\omega\in\mathbb{R}, \theta\in[0,2\pi]$,
\begin{eqnarray} \label{singleGFs}
\G_{0}(\omega,\theta)  &\equiv& (\omega e^{i\theta} - \NH_{0})^{-1},
\end{eqnarray}
and define its complex conjugate, $\G_{0}^{\dagger}(\omega,\text{-}\theta)$, parameterized by $\text{-}\theta$, see also Supplemental Material \ref{GFcorrappendix}. This is subtle because $\GG_{0}$ is only defined for $\omega \in\mathbb{R}$, and the zeros of $\G_{0},\G_{0}^{\dagger}$ are, in general, complex. However, we note that $\omega^{2} = (\omega e^{i\theta})(\omega e^{\text{-}i\theta})$ for any $\theta\in[0,2\pi]$ and, hence, are free to choose $\theta$ as function of $\omega$, defining a path in the complex plane. We can then factor $\GG_{0}$ (Supplemental Material \ref{GFcorrappendix}) as
\begin{eqnarray}
\label{baredGF}
 \GG_{0} =	
\begin{pmatrix}
0 & \G_{0}^{\dagger} 
\\ \G_{0} & 0
\end{pmatrix}  \left[ 1 -
\omega \begin{pmatrix}
0 & \G_{0}^{\dagger}\\ \G_{0}  & 0	
\end{pmatrix}
\begin{pmatrix}
\text{-}1 &  e^{i\theta} \\  e^{\text{-}i\theta}  & \text{-}1	
\end{pmatrix} 
\right]^{-1}.
\end{eqnarray}
\no  We choose $\theta(\omega)$ such that our path intersects the zeros, $\omega_{*},\theta_{*}$ of $\G_{0}$. To sum up, the doubled band gap is defined by the radial distance, $\omega$, between bands as shown in Fig.~\ref{figure1}, and ``in-gap" zeros of the doubled Green's function correspond to ``in-gap" zeros of the single non-Hermitian Green's functions.\\

\begin{table}[]
\figuretitle{Non-Hermitian Boundary Modes}
    \centering
    \begin{tabular}{ |l| l | l |l|l | p{5cm} |}
    \hline
    Case & sGF & dGF & Gap & Manifestation\\ \hline
    I & $\theta = 0,\pi$  & Yes & Any & Traditional Bulk-Boundary \\ \hline
    II & $\theta \neq 0,\pi$  & Yes  & Any & Anomalous Skin Effect \\  \hline
    III & Singular  & Yes  & Point & Trivial Skin Effect\\  \hline
    IV & N/A & No & Point & Gapless \\ \hline
    \end{tabular}
    \caption{Possible correspondences of single vs. doubled Green's function zeros (sGF vs dGF) by Eq. \eqref{baredGF}, with $\lbrace\theta = \arg(\omega),\ \omega\in\mathbb{C}\rbrace$. Doubled Green's functions defined with respect to specified gap condition. For topologically non-trivial bulk, topological edge modes are present in Cases I, II. Edge modes in Case III are not topological. Case IV equivalent to Hermitian gapless phase.}
    \label{tabulate}
\end{table}
%%%%%%%%%%%%%%%%%%%%%%%%%%%%%%%%%%%%%%%%%%%%%%%%%%%%%%%%%%%%%%%%
\indent \textit{Non-Hermitian Boundary Modes }- Projecting the Green's functions to an edge, we come to a main result, the exhaustive determination of the possible boundary modes in non-Hermitian systems, Table \ref{tabulate}. Namely, We consider the consequences of Eq. \eqref{baredGF} for different gap conditions. \\
\indent First, in the presence of a line gap (Fig.~\ref{figure1} a), if the projected doubled Green's function, $\GG_{0}(\omega,\vv k_{\parallel},\vv r_{\perp} = 0)$, has in-gap zeros, we distinguish two topological cases.  Either, the projected single Green's function, $\text{G}_{0}(\omega,\vv k_{\parallel},\vv r_{\perp} = 0)$, has zeros on the real axis, $\theta_{*} = 0,\pi$ (I),  or in the complex plane, $\theta_{*}\neq0,\pi$ (II). Case I corresponds to \textit{traditional bulk boundary correspondence} -- a topologically non-trivial phase hosts topological edge modes reflecting the bulk topological degeneracy. Case II we refer to as the \textit{Anomalous Skin Effect} -- a topologically non-trivial phase hosts topological edge modes with complex conjugate energies, a growing and a decaying mode. Since band topology is defined by the doubled Hamiltonian, cases I and II are not topologically distinct and generalize the notion of bulk boundary correspondence.\\
\indent Next, consider a single band point gap (Fig.~\ref{figure1}b). Edges cut the band and project along corresponding momenta, with two possible outcomes. The point gap is either preserved, or the bands are ``flattened" -- contractible to a point -- and the point gap topology trivialized, see also Supplemental Material \ref{appendPoint}). By contrast, the doubled Green's function topology is defined over the (real) doubled gap before projection (Fig.~\ref{figure1}b) in both cases. If the point gap survives, the single projected Green's function is non-singular. Hence, non-trivial bulk point gap topology has corresponding boundary modes by Eq. \eqref{baredGF}, case I and II. Alternatively, if bands are flattened, the projected spectrum is gapless and the single projected Green's function is singular. A bulk point gap invariant does not have corresponding topological edge modes. Instead, we observe a \textit{Trivial Skin Effect} (case III) -- topologically trivial localization of bulk modes at the edge. Heuristically, if an edge destroys the entire bulk topology, it must carry all bulk topological information and hence localize all bulk modes, see also Supplemental Material \ref{appendPoint}.

\indent Now, consider a topologically trivial line gap topology, an absence of in-gap zeros for the doubled Green's function. Bands in the complex plane generically form closed loops and thus, also have point gaps, e.g. band centers in Fig.~\ref{figure1}a. The point gap topology is always trivial with respect to the line gap. However, if the point gap invariant is intrinsically non-trivial and an edge flattens the bands, a trivial skin effect (III) is generated as before. Skin modes are not topological and have energies away from the line gap, lying in the bands, but can be detected by computing the double Green's function of a single band with respect to its center. Since a band may have both a point gap and line gap invariant, the trivial skin effect may coexist with topological edge-localized modes (I,II). Note that for line gaps, $\omega =0$ modes are special in that they are always in-gap and energetically separated from the trivial skin modes \cite{PhysRevLett.121.136802}. 

\indent Finally, consider a band gap such as in Fig.~\ref{figure1}c. Here the doubled Green's function is directly sensitive to both the band and point gap topology, and the zero crossings are, by definition, separated in energy (Fig.~\ref{figure1}c). If the point gap is trivialized, we observe a trivial skin effect, otherwise, we see topological edge modes corresponding to the point gap topology. On top of these modes, the topological edge modes corresponding to the band gap topology can be detected via the same correspondence in Eq. \ref{baredGF}, and Table \ref{tabulate} holds, as above.\\

\indent \textit{The Role of Symmetry} - Since cases I and II are not topologically distinct within the doubled Hamiltonian classification, traditional bulk-boundary correspondence must be a symmetry protected property. Instead, Eq. \eqref{baredGF} provides a purely topological generalization of bulk boundary correspondence, unifying cases I and II, protected by the existence of a gap, see also Supplemental Material \ref{GFcorrappendix}. In this sense, traditional bulk-boundary correspondence is a special case (choosing the real axis) of the anomalous skin effect. In fact, the spectrum can be continuously rotated in Fig.~\ref{figure1}c, tuning between cases I and II. Furthermore, while a point gap does not protect bulk boundary correspondence, bulk vs. edge symmetries still determine the emergence of the trivial skin effect. This suggests a possible classification, via the above formalism, of non-Hermitian SPTs \cite{2018arXiv181210490Z,kawabata2019topological,kawabata2018symmetry} by their preservation of traditional bulk-boundary correspondence and/or by the absence of a trivial skin effect.\\

\indent \textit{Distinguishing Topological Invariants} - Characterizing system edge modes, the formalism makes it possible probe non-Hermitian band topology directly. For example, consider $\mathbb{Z}$ and $\mathbb{Z}_{2}$ topological invariants.  The edge dispersion parity under time-reversal symmetry (TRS) immediately distinguishes them. This distinction is critical in non-Hermitian systems, where topological invariants depend on gap conditions. The doubled Hamiltonian is sensitive to gap conditions \cite{kawabata2019topological,kawabata2018symmetry,2018arXiv181210490Z}, and, by extension, the doubled Green's function detects ``in-gap" edge localized modes for any gap topology, see Supplemental Material \ref{bandappend}. In fact, in the example below, we identify a purely non-Hermitian topological phase transition, where line gap separated bands are tuned to a point gap, without altering Hamiltonian symmetries, Fig.~\ref{Chern}.\\
%%%%%%%%%%%%%%%%%%%%%%%%%%%%%%%%%%%%%%%%%%%
\begin{figure*}
\begin{center}
\figuretitle{Non-Hermitian Chern Insulator Boundary Modes}
\includegraphics[width = 1 \textwidth]{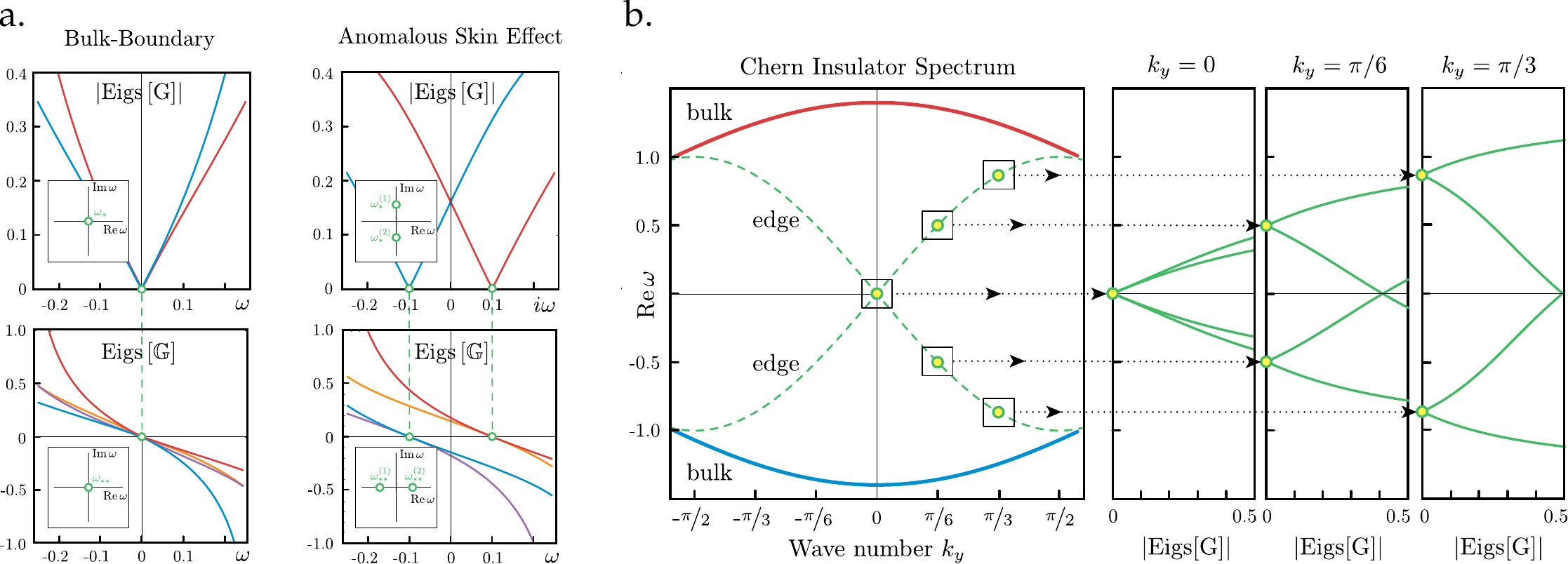}
%\vspace{5 mm}
%\includegraphics[width = 0.48 \textwidth]{EdgeStates_no_title.pdf}	
\end{center}
	\caption{Bulk boundary correspondences in non-Hermitian Chern Insulator. Panel a. Top (Bottom) row corresponds to single (doubled) Green's function. System exhibits both types of topological boundary modes: Case I, left column, e.g. $m=0.4, k_{y} = 0$, $h_x=h_y=0.1$, $h_z=0$, and
	Case II, right column, e.g. $m=0.4, k_{y} = 0$, $h_x=h_y=0$, $h_z=0.1$. 
	Panel b. Edge state dispersion ($h_{z} = 0, m = 0.6$, edge along $\Hat{y}$-axis). For every in-plane momentum $\k_\parallel$, each $k_{y}$ above, we solve for boundary mode energy, defining a dispersion relation.}
\label{edge}
\end{figure*}
%%%%%%%%%%%%%%%%%%%%%%%%%%%%%%%%%%%%%%%%%%%

\indent \textit{Examples} - We illustrate the value of our universal framework in the context of the \textit{non-Hermitian Chern Insulator}. We also checked that our formalism works for edges of arbitrary co-dimensions, see also \cite{Slager2015}, and for the well studied non-Hermitian version of the {\it Su-Schriefffer-Heeger (SSH) model} \cite{RevModPhys.60.781}, see Supplemental Material \ref{sshappend}.\\
\indent Specifically, we consider the Hamiltonian,
%%%%%%%%%%%%%%%%%%%%%%%%%%%%%%%%%%%%%%%%%%%
\begin{align}
	\label{Hamiltonian}
	\mathcal{H} =
	\vv \xi  \, \vv \sigma 
	+ i \vec h \vv \sigma,
\end{align}
%%%%%%%%%%%%%%%%%%%%%%%%%%%%%%%%%%%%%%%%%%%%%
\no where $\vv\sigma = (\sigma_x,\sigma_y, \sigma_z)$,  $\vv h  = (h_x, h_y, h_z)$ indicates the strength of the non-Hermitian field, and $\vv \xi  = (\cos k_x+\cos k_{y} -m, -\sin k_x, \sin k_{y})$. We use the methods in \cite{Slager2015} to compute the single and doubled Green's functions zeros for this model, see Supplemental Material \ref{chernpolesappend}.\\
%%%%%%%%%%%%%%%%%%%%%%%%%%%%%%%%%%%%%%%%%%%%%%%
\begin{figure*}
\figuretitle{Non-Hermitian Chern Insulator Phase Diagram}
	\includegraphics[width = .95 \textwidth]{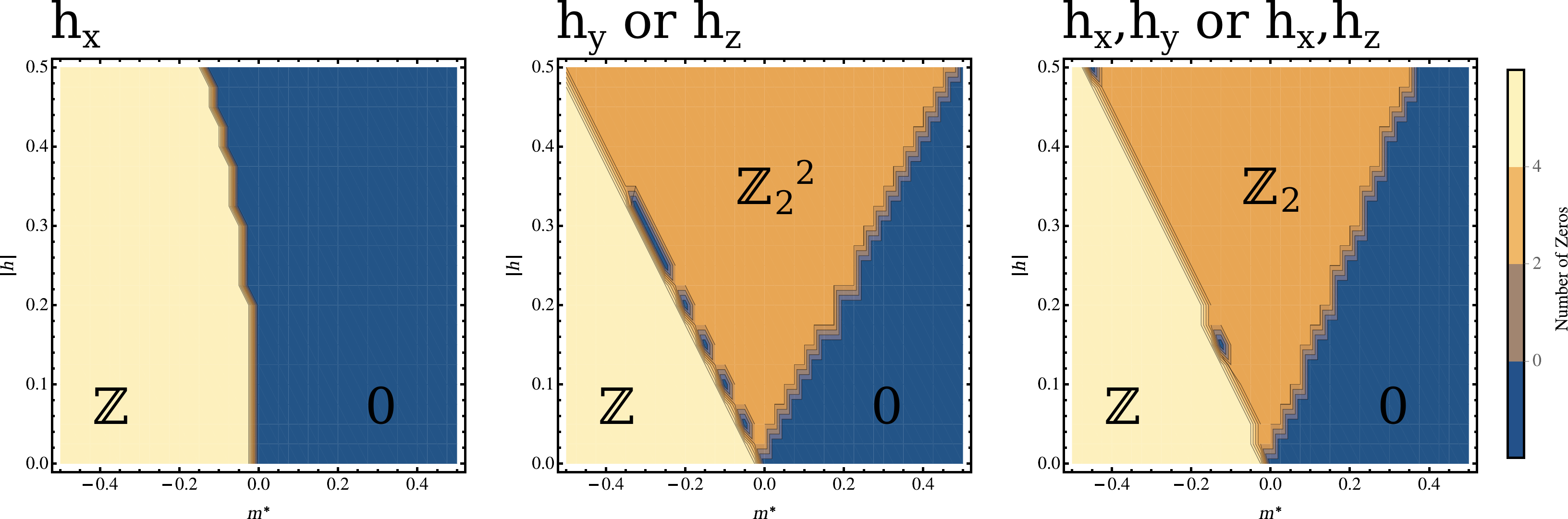}
\caption{Phases of non-Hermitian Chern insulator. Non-zero components of non-Hermitian field, $\vv h$, are indicated in figure panels. Transition between 4 and 2 zeros signals line to point gap transition, topological invariants labeled. Note, absence of non-trivial point gap when only $h_{x}\neq 0$. Components of $\vv h$ chosen to be anisotropic. Shown are representative cases for given topological invariants}
	\label{Chern} 
\end{figure*}
%%%%%%%%%%%%%%%%%%%%%%%%%%%%%%%%%%%%%%%%%%%%%%%%%
\indent  In general, the dispersion relation can be read off from the Green's function zeros. Here, projecting the Hamiltonian to the edge $\hat{x} = 0$, reduces it to the non-Hermitian SSH model, $\vv \xi = (m-\cos(k_{y}),0,\sin(k_{y}))$. Hence, the edge dispersion is simply given by $\pm[ \sin(k_{y})+ih_{z}]$ (Supplemental Material \ref{chernpolesappend}) and is real for $h_{z} = 0$ by a combination of transposition and chiral symmetry \cite{hirsbrunner2019topology,PhysRevLett.121.136802,Kawabata2018}, see Supplemental Material \ref{sshappend}. Therefore, we observe a generalized bulk-boundary correspondence, case I for $h_{z} = 0$ and case II for $h_{z} \neq0$, see Fig.~\ref{edge}. 

\indent Given the chiral symmetry of this 2D model, we compute the Chern number via Green's function zeros, see Supplemental Material \ref{chernphaseappend}. While the line gap is well defined, counting single Green's Function zeros suffices; two (no) zeros between bands imply $\mathcal{C}_{1} = \pm 1\ \ ($0$)$, see Fig.~\ref{Chern} and Supplemental Material \ref{chernphaseappend}. And, the dispersion is odd under TRS (Fig.~\ref{edge} b), consistent with a $\mathbb{Z}$ invariant. This computation is analytically tractable for simple models. The topological transition between phases is marked by a gap-less region, and, as seen by \cite{PhysRevLett.121.136802}, there exists a regime in which the Bloch Chern number is not defined, $m_{*}\pm \abs{\vv h}$. Here, the bands are ``inseparable" \cite{Shen2018} in the complex plane. This regime is characterized by a point gap invariant.

\indent We examine the topology of the single remaining band via the doubled Green's function. Applying the same zero counting argument as above, we see a new uniquely non-Hermitian transition, Fig.~\ref{Chern}. For $h_{y} = h_{z} = 0$, the Hermitian $m = 2$ phase transition is slightly modified. However, when $h_{y}\neq 0$ or $h_{z} \neq 0$, we see two zeros appear instead of four. We also notice that the doubled dispersion relation in this regime becomes time-reversal symmetric, indicating the bands are indexed by a $\mathbb{Z}_{2}$ invariant instead of a $\mathbb{Z}$ invariant. Note, TRS in non-Hermitian systems corresponds to $\vv k\rightarrow -\vv k$ and transposition. Hence, the spectrum is still odd with respect to $\vv k$, but left and right modes are swapped, implying TRS.\\
\indent This transition is thoroughly understood via all relevant symmetries in Supplemental Material \ref{chernsymappend}. Here, we report the point gap classification of the model in the particular cases of generic $\vv h$ and the special cases $h_{z} = h_{x} = 0$, $h_{z} = h_{y} = 0$, and $h_{x} = h_{y} = 0$. The point gap classification predicts a $\mathbb{Z}_{2}$ invariant for generic $\vv h$, a $\mathbb{Z}_{2}^{2}$ for only $h_{y} \neq 0$ or $h_{z} \neq 0$, and a trivial phase for only $h_{x}\neq 0$ \cite{kawabata2018symmetry,2018arXiv181210490Z}. By contrast, the line gap classification is $\mathbb{Z}$ for small perturbations in all cases \cite{kawabata2018symmetry,2018arXiv181210490Z,PhysRevB.98.024310,chen2018hall}. As we increase $\abs{\vv h}$, the line gap (Fig.~\ref{figure1}a) closes and forms a point gap (Fig.~\ref{figure1}b). While existing methods were unable to provide a direct computation of the topological invariant in this regime \cite{PhysRevLett.121.136802, porras2018topological}, our formalism, counting zeros, is sensitive to these transitions, see Fig.~\ref{Chern}. In fact, comparing our classification of the point gap regime to previous work on the non-Hermitian skin effect for this model, \cite{PhysRevLett.121.136802,Kawabata2018}, we find trivial skin modes emerge for the same conditions as predicted by the edge-induced trivialization of non-trivial point gap topology, $h_{y}\neq 0$.\\

%%%%%%%%%%%%%%%%%%%%%%%%%%%%%%%%%%%%%%%%%%%%%%%%%%%%%%%%%%%%%%%%%
\indent\textit{Discussion and Conclusion} - We presented a universal framework to determine boundary modes in non-Hermitian systems with two major outcomes, see Table \ref{tabulate}. The first is a generalization of bulk-boundary correspondence to non-Hermitian systems in the presence of a band gap. We distinguish two types of topological edge modes, those obeying traditional bulk-boundary correspondence (I) and the anomalous skin effect (II), demonstrating traditional bulk-boundary correspondence to be a symmetry constraint on the spectrum. This is accomplished via a complete characterization of topological edge modes and their dispersion relations, allowing us to detect and distinguish different topological bulk invariants.\\  
\indent Second, our framework detects a uniquely non-Hermitian phase transition under the closing of a line gap (Fig.~\ref{figure1} a) into a point gap (Fig.~\ref{figure1} b). As presented in Table \ref{tabulate}, non-trivial point gap topology does not guarantee bulk-boundary correspondence. In particular an edge may either trivialize or preserve the bulk point gap invariant. We explain the trivial non-Hermitian skin effect as this trivialization (case III), a break down of bulk boundary correspondence. And, we detect topological edge modes associated with a bulk point gap in the latter case (I,II). Thus, our framework makes novel non-Hermitian phase transitions physically accessible. This suggests the extended SPT classification \cite{2018arXiv181210490Z,kawabata2018symmetry} is relevant beyond fine tuning and indicates the existence of physically accessible intermediate topological phases in generic open systems. We illustrated this by computing the phase diagram of the non-Hermitian Chern insulator under open boundary conditions, Fig.~\ref{Chern}, for a gapless parameter regime inaccessible to previous methods \cite{PhysRevLett.121.136802}.\\

\noindent\textit{Acknowledgements} - The authors would like to especially thank Jong Yeon Lee for his suggestion to inspect boundary modes of the doubled Hamiltonian and for helpful discussions on the physical significance of edge modes in the presence of a point gap. We also thank Hengyun Zhou for helpful insights on the classification scheme in \cite{2018arXiv181210490Z}. We cordially thank Ashvin Vishwanath, Joaquin Rodriguez Nieva, Ching Hua Lee for valuable discussions. A.J.K. was  supported  by  the  Swiss  National  Science  Foundation, grant  P2ELP2\_175278. R.-J.S appreciatively acknowledges funding via Ashvin Vishwanath from the Center for the Advancement of Topological Materials initiative, an Energy Frontier Research Center funded by the U.S. Department of Energy, Office of Science. 

\bibliography{Refs}

%merlin.mbs apsrev4-1.bst 2010-07-25 4.21a (PWD, AO, DPC) hacked
%Control: key (0)
%Control: author (8) initials jnrlst
%Control: editor formatted (1) identically to author
%Control: production of article title (-1) disabled
%Control: page (0) single
%Control: year (1) truncated
%Control: production of eprint (0) enabled
\begin{thebibliography}{53}%
\makeatletter
\providecommand \@ifxundefined [1]{%
 \@ifx{#1\undefined}
}%
\providecommand \@ifnum [1]{%
 \ifnum #1\expandafter \@firstoftwo
 \else \expandafter \@secondoftwo
 \fi
}%
\providecommand \@ifx [1]{%
 \ifx #1\expandafter \@firstoftwo
 \else \expandafter \@secondoftwo
 \fi
}%
\providecommand \natexlab [1]{#1}%
\providecommand \enquote  [1]{``#1''}%
\providecommand \bibnamefont  [1]{#1}%
\providecommand \bibfnamefont [1]{#1}%
\providecommand \citenamefont [1]{#1}%
\providecommand \href@noop [0]{\@secondoftwo}%
\providecommand \href [0]{\begingroup \@sanitize@url \@href}%
\providecommand \@href[1]{\@@startlink{#1}\@@href}%
\providecommand \@@href[1]{\endgroup#1\@@endlink}%
\providecommand \@sanitize@url [0]{\catcode `\\12\catcode `\$12\catcode
  `\&12\catcode `\#12\catcode `\^12\catcode `\_12\catcode `\%12\relax}%
\providecommand \@@startlink[1]{}%
\providecommand \@@endlink[0]{}%
\providecommand \url  [0]{\begingroup\@sanitize@url \@url }%
\providecommand \@url [1]{\endgroup\@href {#1}{\urlprefix }}%
\providecommand \urlprefix  [0]{URL }%
\providecommand \Eprint [0]{\href }%
\providecommand \doibase [0]{http://dx.doi.org/}%
\providecommand \selectlanguage [0]{\@gobble}%
\providecommand \bibinfo  [0]{\@secondoftwo}%
\providecommand \bibfield  [0]{\@secondoftwo}%
\providecommand \translation [1]{[#1]}%
\providecommand \BibitemOpen [0]{}%
\providecommand \bibitemStop [0]{}%
\providecommand \bibitemNoStop [0]{.\EOS\space}%
\providecommand \EOS [0]{\spacefactor3000\relax}%
\providecommand \BibitemShut  [1]{\csname bibitem#1\endcsname}%
\let\auto@bib@innerbib\@empty
%</preamble>
\bibitem [{\citenamefont {{Zhou}}\ and\ \citenamefont
  {{Lee}}(2018)}]{2018arXiv181210490Z}%
  \BibitemOpen
  \bibfield  {author} {\bibinfo {author} {\bibfnamefont {H.}~\bibnamefont
  {{Zhou}}}\ and\ \bibinfo {author} {\bibfnamefont {J.~Y.}\ \bibnamefont
  {{Lee}}},\ }\href@noop {} {\bibfield  {journal} {\bibinfo  {journal} {arXiv
  e-prints}\ ,\ \bibinfo {eid} {arXiv:1812.10490}} (\bibinfo {year} {2018})},\
  \Eprint {http://arxiv.org/abs/1812.10490} {arXiv:1812.10490
  [cond-mat.mes-hall]} \BibitemShut {NoStop}%
\bibitem [{\citenamefont {Kawabata}\ \emph
  {et~al.}(2018{\natexlab{a}})\citenamefont {Kawabata}, \citenamefont
  {Shiozaki},\ and\ \citenamefont {Ueda}}]{Kawabata2018}%
  \BibitemOpen
  \bibfield  {author} {\bibinfo {author} {\bibfnamefont {K.}~\bibnamefont
  {Kawabata}}, \bibinfo {author} {\bibfnamefont {K.}~\bibnamefont {Shiozaki}},
  \ and\ \bibinfo {author} {\bibfnamefont {M.}~\bibnamefont {Ueda}},\
  }\href@noop {} {\bibfield  {journal} {\bibinfo  {journal} {Physical Review
  B}\ }\textbf {\bibinfo {volume} {98}},\ \bibinfo {pages} {165148} (\bibinfo
  {year} {2018}{\natexlab{a}})}\BibitemShut {NoStop}%
\bibitem [{\citenamefont {Kawabata}\ \emph {et~al.}(2019)\citenamefont
  {Kawabata}, \citenamefont {Higashikawa}, \citenamefont {Gong}, \citenamefont
  {Ashida},\ and\ \citenamefont {Ueda}}]{kawabata2019topological}%
  \BibitemOpen
  \bibfield  {author} {\bibinfo {author} {\bibfnamefont {K.}~\bibnamefont
  {Kawabata}}, \bibinfo {author} {\bibfnamefont {S.}~\bibnamefont
  {Higashikawa}}, \bibinfo {author} {\bibfnamefont {Z.}~\bibnamefont {Gong}},
  \bibinfo {author} {\bibfnamefont {Y.}~\bibnamefont {Ashida}}, \ and\ \bibinfo
  {author} {\bibfnamefont {M.}~\bibnamefont {Ueda}},\ }\href@noop {} {\bibfield
   {journal} {\bibinfo  {journal} {Nature Communications}\ }\textbf {\bibinfo
  {volume} {10}},\ \bibinfo {pages} {297} (\bibinfo {year} {2019})}\BibitemShut
  {NoStop}%
\bibitem [{\citenamefont {Kawabata}\ \emph
  {et~al.}(2018{\natexlab{b}})\citenamefont {Kawabata}, \citenamefont
  {Shiozaki}, \citenamefont {Ueda},\ and\ \citenamefont
  {Sato}}]{kawabata2018symmetry}%
  \BibitemOpen
  \bibfield  {author} {\bibinfo {author} {\bibfnamefont {K.}~\bibnamefont
  {Kawabata}}, \bibinfo {author} {\bibfnamefont {K.}~\bibnamefont {Shiozaki}},
  \bibinfo {author} {\bibfnamefont {M.}~\bibnamefont {Ueda}}, \ and\ \bibinfo
  {author} {\bibfnamefont {M.}~\bibnamefont {Sato}},\ }\href@noop {} {\bibfield
   {journal} {\bibinfo  {journal} {arXiv preprint arXiv:1812.09133}\ }
  (\bibinfo {year} {2018}{\natexlab{b}})}\BibitemShut {NoStop}%
\bibitem [{\citenamefont {Leykam}\ \emph {et~al.}(2017)\citenamefont {Leykam},
  \citenamefont {Bliokh}, \citenamefont {Huang}, \citenamefont {Chong},\ and\
  \citenamefont {Nori}}]{leykam2017edge}%
  \BibitemOpen
  \bibfield  {author} {\bibinfo {author} {\bibfnamefont {D.}~\bibnamefont
  {Leykam}}, \bibinfo {author} {\bibfnamefont {K.~Y.}\ \bibnamefont {Bliokh}},
  \bibinfo {author} {\bibfnamefont {C.}~\bibnamefont {Huang}}, \bibinfo
  {author} {\bibfnamefont {Y.~D.}\ \bibnamefont {Chong}}, \ and\ \bibinfo
  {author} {\bibfnamefont {F.}~\bibnamefont {Nori}},\ }\href@noop {} {\bibfield
   {journal} {\bibinfo  {journal} {Physical review letters}\ }\textbf {\bibinfo
  {volume} {118}},\ \bibinfo {pages} {040401} (\bibinfo {year}
  {2017})}\BibitemShut {NoStop}%
\bibitem [{\citenamefont {Shen}\ \emph
  {et~al.}(2018{\natexlab{a}})\citenamefont {Shen}, \citenamefont {Zhen},\ and\
  \citenamefont {Fu}}]{PhysRevLett.120.146402}%
  \BibitemOpen
  \bibfield  {author} {\bibinfo {author} {\bibfnamefont {H.}~\bibnamefont
  {Shen}}, \bibinfo {author} {\bibfnamefont {B.}~\bibnamefont {Zhen}}, \ and\
  \bibinfo {author} {\bibfnamefont {L.}~\bibnamefont {Fu}},\ }\href {\doibase
  10.1103/PhysRevLett.120.146402} {\bibfield  {journal} {\bibinfo  {journal}
  {Phys. Rev. Lett.}\ }\textbf {\bibinfo {volume} {120}},\ \bibinfo {pages}
  {146402} (\bibinfo {year} {2018}{\natexlab{a}})}\BibitemShut {NoStop}%
\bibitem [{\citenamefont {Yao}\ and\ \citenamefont {Wang}(2018)}]{Yao2018}%
  \BibitemOpen
  \bibfield  {author} {\bibinfo {author} {\bibfnamefont {S.}~\bibnamefont
  {Yao}}\ and\ \bibinfo {author} {\bibfnamefont {Z.}~\bibnamefont {Wang}},\
  }\href@noop {} {\bibfield  {journal} {\bibinfo  {journal} {arXiv preprint
  arXiv:1803.01876}\ } (\bibinfo {year} {2018})}\BibitemShut {NoStop}%
\bibitem [{\citenamefont {Esaki}\ \emph {et~al.}(2011)\citenamefont {Esaki},
  \citenamefont {Sato}, \citenamefont {Hasebe},\ and\ \citenamefont
  {Kohmoto}}]{PhysRevB.84.205128}%
  \BibitemOpen
  \bibfield  {author} {\bibinfo {author} {\bibfnamefont {K.}~\bibnamefont
  {Esaki}}, \bibinfo {author} {\bibfnamefont {M.}~\bibnamefont {Sato}},
  \bibinfo {author} {\bibfnamefont {K.}~\bibnamefont {Hasebe}}, \ and\ \bibinfo
  {author} {\bibfnamefont {M.}~\bibnamefont {Kohmoto}},\ }\href {\doibase
  10.1103/PhysRevB.84.205128} {\bibfield  {journal} {\bibinfo  {journal} {Phys.
  Rev. B}\ }\textbf {\bibinfo {volume} {84}},\ \bibinfo {pages} {205128}
  (\bibinfo {year} {2011})}\BibitemShut {NoStop}%
\bibitem [{\citenamefont {Lee}\ \emph {et~al.}(2018)\citenamefont {Lee},
  \citenamefont {Li}, \citenamefont {Liu}, \citenamefont {Tai}, \citenamefont
  {Thomale},\ and\ \citenamefont {Zhang}}]{lee2018tidal}%
  \BibitemOpen
  \bibfield  {author} {\bibinfo {author} {\bibfnamefont {C.~H.}\ \bibnamefont
  {Lee}}, \bibinfo {author} {\bibfnamefont {G.}~\bibnamefont {Li}}, \bibinfo
  {author} {\bibfnamefont {Y.}~\bibnamefont {Liu}}, \bibinfo {author}
  {\bibfnamefont {T.}~\bibnamefont {Tai}}, \bibinfo {author} {\bibfnamefont
  {R.}~\bibnamefont {Thomale}}, \ and\ \bibinfo {author} {\bibfnamefont
  {X.}~\bibnamefont {Zhang}},\ }\href@noop {} {\bibfield  {journal} {\bibinfo
  {journal} {arXiv preprint arXiv:1812.02011}\ } (\bibinfo {year}
  {2018})}\BibitemShut {NoStop}%
\bibitem [{\citenamefont {El-Ganainy}\ \emph {et~al.}(2018)\citenamefont
  {El-Ganainy}, \citenamefont {Makris}, \citenamefont {Khajavikhan},
  \citenamefont {Musslimani}, \citenamefont {Rotter},\ and\ \citenamefont
  {Christodoulides}}]{Ganainy2018}%
  \BibitemOpen
  \bibfield  {author} {\bibinfo {author} {\bibfnamefont {R.}~\bibnamefont
  {El-Ganainy}}, \bibinfo {author} {\bibfnamefont {K.~G.}\ \bibnamefont
  {Makris}}, \bibinfo {author} {\bibfnamefont {M.}~\bibnamefont {Khajavikhan}},
  \bibinfo {author} {\bibfnamefont {Z.~H.}\ \bibnamefont {Musslimani}},
  \bibinfo {author} {\bibfnamefont {S.}~\bibnamefont {Rotter}}, \ and\ \bibinfo
  {author} {\bibfnamefont {D.~N.}\ \bibnamefont {Christodoulides}},\
  }\href@noop {} {\bibfield  {journal} {\bibinfo  {journal} {Nature Physics}\
  }\textbf {\bibinfo {volume} {14}},\ \bibinfo {pages} {11} (\bibinfo {year}
  {2018})}\BibitemShut {NoStop}%
\bibitem [{\citenamefont {Guo}\ \emph {et~al.}(2009)\citenamefont {Guo},
  \citenamefont {Salamo}, \citenamefont {Duchesne}, \citenamefont {Morandotti},
  \citenamefont {Volatier-Ravat}, \citenamefont {Aimez}, \citenamefont
  {Siviloglou},\ and\ \citenamefont {Christodoulides}}]{Guo2009}%
  \BibitemOpen
  \bibfield  {author} {\bibinfo {author} {\bibfnamefont {A.}~\bibnamefont
  {Guo}}, \bibinfo {author} {\bibfnamefont {G.~J.}\ \bibnamefont {Salamo}},
  \bibinfo {author} {\bibfnamefont {D.}~\bibnamefont {Duchesne}}, \bibinfo
  {author} {\bibfnamefont {R.}~\bibnamefont {Morandotti}}, \bibinfo {author}
  {\bibfnamefont {M.}~\bibnamefont {Volatier-Ravat}}, \bibinfo {author}
  {\bibfnamefont {V.}~\bibnamefont {Aimez}}, \bibinfo {author} {\bibfnamefont
  {G.~A.}\ \bibnamefont {Siviloglou}}, \ and\ \bibinfo {author} {\bibfnamefont
  {D.~N.}\ \bibnamefont {Christodoulides}},\ }\href {\doibase
  10.1103/PhysRevLett.103.093902} {\bibfield  {journal} {\bibinfo  {journal}
  {Phys. Rev. Lett.}\ }\textbf {\bibinfo {volume} {103}},\ \bibinfo {pages}
  {093902} (\bibinfo {year} {2009})}\BibitemShut {NoStop}%
\bibitem [{\citenamefont {R{\"u}ter}\ \emph {et~al.}(2010)\citenamefont
  {R{\"u}ter}, \citenamefont {Makris}, \citenamefont {El-Ganainy},
  \citenamefont {Christodoulides}, \citenamefont {Segev},\ and\ \citenamefont
  {Kip}}]{Ruter2010}%
  \BibitemOpen
  \bibfield  {author} {\bibinfo {author} {\bibfnamefont {C.~E.}\ \bibnamefont
  {R{\"u}ter}}, \bibinfo {author} {\bibfnamefont {K.~G.}\ \bibnamefont
  {Makris}}, \bibinfo {author} {\bibfnamefont {R.}~\bibnamefont {El-Ganainy}},
  \bibinfo {author} {\bibfnamefont {D.~N.}\ \bibnamefont {Christodoulides}},
  \bibinfo {author} {\bibfnamefont {M.}~\bibnamefont {Segev}}, \ and\ \bibinfo
  {author} {\bibfnamefont {D.}~\bibnamefont {Kip}},\ }\href@noop {} {\bibfield
  {journal} {\bibinfo  {journal} {Nature physics}\ }\textbf {\bibinfo {volume}
  {6}},\ \bibinfo {pages} {192} (\bibinfo {year} {2010})}\BibitemShut {NoStop}%
\bibitem [{\citenamefont {Zeuner}\ \emph {et~al.}(2015)\citenamefont {Zeuner},
  \citenamefont {Rechtsman}, \citenamefont {Plotnik}, \citenamefont {Lumer},
  \citenamefont {Nolte}, \citenamefont {Rudner}, \citenamefont {Segev},\ and\
  \citenamefont {Szameit}}]{Zeuner2015}%
  \BibitemOpen
  \bibfield  {author} {\bibinfo {author} {\bibfnamefont {J.~M.}\ \bibnamefont
  {Zeuner}}, \bibinfo {author} {\bibfnamefont {M.~C.}\ \bibnamefont
  {Rechtsman}}, \bibinfo {author} {\bibfnamefont {Y.}~\bibnamefont {Plotnik}},
  \bibinfo {author} {\bibfnamefont {Y.}~\bibnamefont {Lumer}}, \bibinfo
  {author} {\bibfnamefont {S.}~\bibnamefont {Nolte}}, \bibinfo {author}
  {\bibfnamefont {M.~S.}\ \bibnamefont {Rudner}}, \bibinfo {author}
  {\bibfnamefont {M.}~\bibnamefont {Segev}}, \ and\ \bibinfo {author}
  {\bibfnamefont {A.}~\bibnamefont {Szameit}},\ }\href {\doibase
  10.1103/PhysRevLett.115.040402} {\bibfield  {journal} {\bibinfo  {journal}
  {Phys. Rev. Lett.}\ }\textbf {\bibinfo {volume} {115}},\ \bibinfo {pages}
  {040402} (\bibinfo {year} {2015})}\BibitemShut {NoStop}%
\bibitem [{\citenamefont {Zhou}\ \emph {et~al.}(2018)\citenamefont {Zhou},
  \citenamefont {Peng}, \citenamefont {Yoon}, \citenamefont {Hsu},
  \citenamefont {Nelson}, \citenamefont {Fu}, \citenamefont {Joannopoulos},
  \citenamefont {Solja{\v c}i{\'c}},\ and\ \citenamefont
  {Zhen}}]{Zhoueaap9859}%
  \BibitemOpen
  \bibfield  {author} {\bibinfo {author} {\bibfnamefont {H.}~\bibnamefont
  {Zhou}}, \bibinfo {author} {\bibfnamefont {C.}~\bibnamefont {Peng}}, \bibinfo
  {author} {\bibfnamefont {Y.}~\bibnamefont {Yoon}}, \bibinfo {author}
  {\bibfnamefont {C.~W.}\ \bibnamefont {Hsu}}, \bibinfo {author} {\bibfnamefont
  {K.~A.}\ \bibnamefont {Nelson}}, \bibinfo {author} {\bibfnamefont
  {L.}~\bibnamefont {Fu}}, \bibinfo {author} {\bibfnamefont {J.~D.}\
  \bibnamefont {Joannopoulos}}, \bibinfo {author} {\bibfnamefont
  {M.}~\bibnamefont {Solja{\v c}i{\'c}}}, \ and\ \bibinfo {author}
  {\bibfnamefont {B.}~\bibnamefont {Zhen}},\ }\href {\doibase
  10.1126/science.aap9859} {\bibfield  {journal} {\bibinfo  {journal}
  {Science}\ } (\bibinfo {year} {2018}),\ 10.1126/science.aap9859}\BibitemShut
  {NoStop}%
\bibitem [{\citenamefont {Shen}\ \emph
  {et~al.}(2018{\natexlab{b}})\citenamefont {Shen}, \citenamefont {Zhen},\ and\
  \citenamefont {Fu}}]{Shen2018}%
  \BibitemOpen
  \bibfield  {author} {\bibinfo {author} {\bibfnamefont {H.}~\bibnamefont
  {Shen}}, \bibinfo {author} {\bibfnamefont {B.}~\bibnamefont {Zhen}}, \ and\
  \bibinfo {author} {\bibfnamefont {L.}~\bibnamefont {Fu}},\ }\href@noop {}
  {\bibfield  {journal} {\bibinfo  {journal} {Physical review letters}\
  }\textbf {\bibinfo {volume} {120}},\ \bibinfo {pages} {146402} (\bibinfo
  {year} {2018}{\natexlab{b}})}\BibitemShut {NoStop}%
\bibitem [{\citenamefont {Gong}\ \emph {et~al.}(2018)\citenamefont {Gong},
  \citenamefont {Ashida}, \citenamefont {Kawabata}, \citenamefont {Takasan},
  \citenamefont {Higashikawa},\ and\ \citenamefont {Ueda}}]{Gong2018}%
  \BibitemOpen
  \bibfield  {author} {\bibinfo {author} {\bibfnamefont {Z.}~\bibnamefont
  {Gong}}, \bibinfo {author} {\bibfnamefont {Y.}~\bibnamefont {Ashida}},
  \bibinfo {author} {\bibfnamefont {K.}~\bibnamefont {Kawabata}}, \bibinfo
  {author} {\bibfnamefont {K.}~\bibnamefont {Takasan}}, \bibinfo {author}
  {\bibfnamefont {S.}~\bibnamefont {Higashikawa}}, \ and\ \bibinfo {author}
  {\bibfnamefont {M.}~\bibnamefont {Ueda}},\ }\href {\doibase
  10.1103/PhysRevX.8.031079} {\bibfield  {journal} {\bibinfo  {journal} {Phys.
  Rev. X}\ }\textbf {\bibinfo {volume} {8}},\ \bibinfo {pages} {031079}
  (\bibinfo {year} {2018})}\BibitemShut {NoStop}%
\bibitem [{\citenamefont {Bender}\ and\ \citenamefont
  {Boettcher}(1998)}]{Bender1998}%
  \BibitemOpen
  \bibfield  {author} {\bibinfo {author} {\bibfnamefont {C.~M.}\ \bibnamefont
  {Bender}}\ and\ \bibinfo {author} {\bibfnamefont {S.}~\bibnamefont
  {Boettcher}},\ }\href@noop {} {\bibfield  {journal} {\bibinfo  {journal}
  {Physical Review Letters}\ }\textbf {\bibinfo {volume} {80}},\ \bibinfo
  {pages} {5243} (\bibinfo {year} {1998})}\BibitemShut {NoStop}%
\bibitem [{\citenamefont {Mostafazadeh}(2002)}]{doi:10.1063/1.1418246}%
  \BibitemOpen
  \bibfield  {author} {\bibinfo {author} {\bibfnamefont {A.}~\bibnamefont
  {Mostafazadeh}},\ }\href {\doibase 10.1063/1.1418246} {\bibfield  {journal}
  {\bibinfo  {journal} {Journal of Mathematical Physics}\ }\textbf {\bibinfo
  {volume} {43}},\ \bibinfo {pages} {205} (\bibinfo {year} {2002})}\BibitemShut
  {NoStop}%
\bibitem [{\citenamefont {Feinberg}\ and\ \citenamefont
  {Zee}(1999)}]{PhysRevE.59.6433}%
  \BibitemOpen
  \bibfield  {author} {\bibinfo {author} {\bibfnamefont {J.}~\bibnamefont
  {Feinberg}}\ and\ \bibinfo {author} {\bibfnamefont {A.}~\bibnamefont {Zee}},\
  }\href {\doibase 10.1103/PhysRevE.59.6433} {\bibfield  {journal} {\bibinfo
  {journal} {Phys. Rev. E}\ }\textbf {\bibinfo {volume} {59}},\ \bibinfo
  {pages} {6433} (\bibinfo {year} {1999})}\BibitemShut {NoStop}%
\bibitem [{\citenamefont {Yao}\ \emph {et~al.}(2018)\citenamefont {Yao},
  \citenamefont {Song},\ and\ \citenamefont {Wang}}]{PhysRevLett.121.136802}%
  \BibitemOpen
  \bibfield  {author} {\bibinfo {author} {\bibfnamefont {S.}~\bibnamefont
  {Yao}}, \bibinfo {author} {\bibfnamefont {F.}~\bibnamefont {Song}}, \ and\
  \bibinfo {author} {\bibfnamefont {Z.}~\bibnamefont {Wang}},\ }\href {\doibase
  10.1103/PhysRevLett.121.136802} {\bibfield  {journal} {\bibinfo  {journal}
  {Phys. Rev. Lett.}\ }\textbf {\bibinfo {volume} {121}},\ \bibinfo {pages}
  {136802} (\bibinfo {year} {2018})}\BibitemShut {NoStop}%
\bibitem [{\citenamefont {Alvarez}\ \emph {et~al.}(2018)\citenamefont
  {Alvarez}, \citenamefont {Vargas}, \citenamefont {Berdakin},\ and\
  \citenamefont {Torres}}]{alvarez2018topological}%
  \BibitemOpen
  \bibfield  {author} {\bibinfo {author} {\bibfnamefont {V.}~\bibnamefont
  {Alvarez}}, \bibinfo {author} {\bibfnamefont {J.}~\bibnamefont {Vargas}},
  \bibinfo {author} {\bibfnamefont {M.}~\bibnamefont {Berdakin}}, \ and\
  \bibinfo {author} {\bibfnamefont {L.}~\bibnamefont {Torres}},\ }\href@noop {}
  {\bibfield  {journal} {\bibinfo  {journal} {arXiv preprint arXiv:1805.08200}\
  } (\bibinfo {year} {2018})}\BibitemShut {NoStop}%
\bibitem [{\citenamefont {Xiong}(2018)}]{xiong2018does}%
  \BibitemOpen
  \bibfield  {author} {\bibinfo {author} {\bibfnamefont {Y.}~\bibnamefont
  {Xiong}},\ }\href@noop {} {\bibfield  {journal} {\bibinfo  {journal} {Journal
  of Physics Communications}\ }\textbf {\bibinfo {volume} {2}},\ \bibinfo
  {pages} {035043} (\bibinfo {year} {2018})}\BibitemShut {NoStop}%
\bibitem [{\citenamefont {Lee}\ and\ \citenamefont
  {Thomale}(2018)}]{lee2018anatomy}%
  \BibitemOpen
  \bibfield  {author} {\bibinfo {author} {\bibfnamefont {C.~H.}\ \bibnamefont
  {Lee}}\ and\ \bibinfo {author} {\bibfnamefont {R.}~\bibnamefont {Thomale}},\
  }\href@noop {} {\bibfield  {journal} {\bibinfo  {journal} {arXiv preprint
  arXiv:1809.02125}\ } (\bibinfo {year} {2018})}\BibitemShut {NoStop}%
\bibitem [{\citenamefont {Kunst}\ and\ \citenamefont
  {Dwivedi}(2018)}]{Kunst2018a}%
  \BibitemOpen
  \bibfield  {author} {\bibinfo {author} {\bibfnamefont {F.~K.}\ \bibnamefont
  {Kunst}}\ and\ \bibinfo {author} {\bibfnamefont {V.}~\bibnamefont
  {Dwivedi}},\ }\href@noop {} {\bibfield  {journal} {\bibinfo  {journal} {arXiv
  preprint arXiv:1812.02186}\ } (\bibinfo {year} {2018})}\BibitemShut {NoStop}%
\bibitem [{\citenamefont {Kunst}\ \emph {et~al.}(2018)\citenamefont {Kunst},
  \citenamefont {Edvardsson}, \citenamefont {Budich},\ and\ \citenamefont
  {Bergholtz}}]{Kunst2018b}%
  \BibitemOpen
  \bibfield  {author} {\bibinfo {author} {\bibfnamefont {F.~K.}\ \bibnamefont
  {Kunst}}, \bibinfo {author} {\bibfnamefont {E.}~\bibnamefont {Edvardsson}},
  \bibinfo {author} {\bibfnamefont {J.~C.}\ \bibnamefont {Budich}}, \ and\
  \bibinfo {author} {\bibfnamefont {E.~J.}\ \bibnamefont {Bergholtz}},\ }\href
  {\doibase 10.1103/PhysRevLett.121.026808} {\bibfield  {journal} {\bibinfo
  {journal} {Phys. Rev. Lett.}\ }\textbf {\bibinfo {volume} {121}},\ \bibinfo
  {pages} {026808} (\bibinfo {year} {2018})}\BibitemShut {NoStop}%
\bibitem [{\citenamefont {{Herviou}}\ \emph {et~al.}(2018)\citenamefont
  {{Herviou}}, \citenamefont {{Bardarson}},\ and\ \citenamefont
  {{Regnault}}}]{2019arXiv190100010H}%
  \BibitemOpen
  \bibfield  {author} {\bibinfo {author} {\bibfnamefont {L.}~\bibnamefont
  {{Herviou}}}, \bibinfo {author} {\bibfnamefont {J.~H.}\ \bibnamefont
  {{Bardarson}}}, \ and\ \bibinfo {author} {\bibfnamefont {N.}~\bibnamefont
  {{Regnault}}},\ }\href@noop {} {\bibfield  {journal} {\bibinfo  {journal}
  {arXiv e-prints}\ ,\ \bibinfo {eid} {arXiv:1901.00010}} (\bibinfo {year}
  {2018})},\ \Eprint {http://arxiv.org/abs/1901.00010} {arXiv:1901.00010
  [cond-mat.mes-hall]} \BibitemShut {NoStop}%
\bibitem [{\citenamefont {Jin}\ and\ \citenamefont {Song}(2018)}]{jin2018bulk}%
  \BibitemOpen
  \bibfield  {author} {\bibinfo {author} {\bibfnamefont {L.}~\bibnamefont
  {Jin}}\ and\ \bibinfo {author} {\bibfnamefont {Z.}~\bibnamefont {Song}},\
  }\href@noop {} {\bibfield  {journal} {\bibinfo  {journal} {arXiv preprint
  arXiv:1809.03139}\ } (\bibinfo {year} {2018})}\BibitemShut {NoStop}%
\bibitem [{\citenamefont {Kitaev}(2009{\natexlab{a}})}]{Clas1a}%
  \BibitemOpen
  \bibfield  {author} {\bibinfo {author} {\bibfnamefont {A.}~\bibnamefont
  {Kitaev}}\ }(\bibinfo {organization} {AIP},\ \bibinfo {year} {2009})\ pp.\
  \bibinfo {pages} {22--30}\BibitemShut {NoStop}%
\bibitem [{\citenamefont {Ryu}\ \emph {et~al.}(2010)\citenamefont {Ryu},
  \citenamefont {Schnyder}, \citenamefont {Furusaki},\ and\ \citenamefont
  {Ludwig}}]{Clas1b}%
  \BibitemOpen
  \bibfield  {author} {\bibinfo {author} {\bibfnamefont {S.}~\bibnamefont
  {Ryu}}, \bibinfo {author} {\bibfnamefont {A.~P.}\ \bibnamefont {Schnyder}},
  \bibinfo {author} {\bibfnamefont {A.}~\bibnamefont {Furusaki}}, \ and\
  \bibinfo {author} {\bibfnamefont {A.~W.}\ \bibnamefont {Ludwig}},\
  }\href@noop {} {\bibfield  {journal} {\bibinfo  {journal} {New Journal of
  Physics}\ }\textbf {\bibinfo {volume} {12}},\ \bibinfo {pages} {065010}
  (\bibinfo {year} {2010})}\BibitemShut {NoStop}%
\bibitem [{\citenamefont {Fu}(2011)}]{Clas1c}%
  \BibitemOpen
  \bibfield  {author} {\bibinfo {author} {\bibfnamefont {L.}~\bibnamefont
  {Fu}},\ }\href {\doibase 10.1103/PhysRevLett.106.106802} {\bibfield
  {journal} {\bibinfo  {journal} {Phys. Rev. Lett.}\ }\textbf {\bibinfo
  {volume} {106}},\ \bibinfo {pages} {106802} (\bibinfo {year}
  {2011})}\BibitemShut {NoStop}%
\bibitem [{\citenamefont {Slager}\ \emph {et~al.}(2012)\citenamefont {Slager},
  \citenamefont {Mesaros}, \citenamefont {Juri{\v c}i{\'c}},\ and\
  \citenamefont {Zaanen}}]{Clas2}%
  \BibitemOpen
  \bibfield  {author} {\bibinfo {author} {\bibfnamefont {R.-J.}\ \bibnamefont
  {Slager}}, \bibinfo {author} {\bibfnamefont {A.}~\bibnamefont {Mesaros}},
  \bibinfo {author} {\bibfnamefont {V.}~\bibnamefont {Juri{\v c}i{\'c}}}, \
  and\ \bibinfo {author} {\bibfnamefont {J.}~\bibnamefont {Zaanen}},\ }\href
  {http://dx.doi.org/10.1038/nphys2513} {\bibfield  {journal} {\bibinfo
  {journal} {Nature Physics}\ }\textbf {\bibinfo {volume} {9}},\ \bibinfo
  {pages} {98} (\bibinfo {year} {2012})}\BibitemShut {NoStop}%
\bibitem [{\citenamefont {Shiozaki}\ and\ \citenamefont
  {Sato}(2014)}]{PhysRevB.90.165114}%
  \BibitemOpen
  \bibfield  {author} {\bibinfo {author} {\bibfnamefont {K.}~\bibnamefont
  {Shiozaki}}\ and\ \bibinfo {author} {\bibfnamefont {M.}~\bibnamefont
  {Sato}},\ }\href {\doibase 10.1103/PhysRevB.90.165114} {\bibfield  {journal}
  {\bibinfo  {journal} {Phys. Rev. B}\ }\textbf {\bibinfo {volume} {90}},\
  \bibinfo {pages} {165114} (\bibinfo {year} {2014})}\BibitemShut {NoStop}%
\bibitem [{\citenamefont {Kruthoff}\ \emph {et~al.}(2017)\citenamefont
  {Kruthoff}, \citenamefont {de~Boer}, \citenamefont {van Wezel}, \citenamefont
  {Kane},\ and\ \citenamefont {Slager}}]{Clas3}%
  \BibitemOpen
  \bibfield  {author} {\bibinfo {author} {\bibfnamefont {J.}~\bibnamefont
  {Kruthoff}}, \bibinfo {author} {\bibfnamefont {J.}~\bibnamefont {de~Boer}},
  \bibinfo {author} {\bibfnamefont {J.}~\bibnamefont {van Wezel}}, \bibinfo
  {author} {\bibfnamefont {C.~L.}\ \bibnamefont {Kane}}, \ and\ \bibinfo
  {author} {\bibfnamefont {R.-J.}\ \bibnamefont {Slager}},\ }\href {\doibase
  10.1103/PhysRevX.7.041069} {\bibfield  {journal} {\bibinfo  {journal} {Phys.
  Rev. X}\ }\textbf {\bibinfo {volume} {7}},\ \bibinfo {pages} {041069}
  (\bibinfo {year} {2017})}\BibitemShut {NoStop}%
\bibitem [{\citenamefont {H\"oller}\ and\ \citenamefont
  {Alexandradinata}(2018)}]{PhysRevB.98.024310}%
  \BibitemOpen
  \bibfield  {author} {\bibinfo {author} {\bibfnamefont {J.}~\bibnamefont
  {H\"oller}}\ and\ \bibinfo {author} {\bibfnamefont {A.}~\bibnamefont
  {Alexandradinata}},\ }\href {\doibase 10.1103/PhysRevB.98.024310} {\bibfield
  {journal} {\bibinfo  {journal} {Phys. Rev. B}\ }\textbf {\bibinfo {volume}
  {98}},\ \bibinfo {pages} {024310} (\bibinfo {year} {2018})}\BibitemShut
  {NoStop}%
\bibitem [{\citenamefont {Po}\ \emph {et~al.}(2017)\citenamefont {Po},
  \citenamefont {Vishwanath},\ and\ \citenamefont {Watanabe}}]{Clas4}%
  \BibitemOpen
  \bibfield  {author} {\bibinfo {author} {\bibfnamefont {H.~C.}\ \bibnamefont
  {Po}}, \bibinfo {author} {\bibfnamefont {A.}~\bibnamefont {Vishwanath}}, \
  and\ \bibinfo {author} {\bibfnamefont {H.}~\bibnamefont {Watanabe}},\ }\href
  {\doibase 10.1038/s41467-017-00133-2} {\bibfield  {journal} {\bibinfo
  {journal} {Nature Communications}\ }\textbf {\bibinfo {volume} {8}},\
  \bibinfo {pages} {50} (\bibinfo {year} {2017})}\BibitemShut {NoStop}%
\bibitem [{\citenamefont {Bradlyn}\ \emph {et~al.}(2017)\citenamefont
  {Bradlyn}, \citenamefont {Elcoro}, \citenamefont {Cano}, \citenamefont
  {Vergniory}, \citenamefont {Wang}, \citenamefont {Felser}, \citenamefont
  {Aroyo},\ and\ \citenamefont {Bernevig}}]{Clas5}%
  \BibitemOpen
  \bibfield  {author} {\bibinfo {author} {\bibfnamefont {B.}~\bibnamefont
  {Bradlyn}}, \bibinfo {author} {\bibfnamefont {L.}~\bibnamefont {Elcoro}},
  \bibinfo {author} {\bibfnamefont {J.}~\bibnamefont {Cano}}, \bibinfo {author}
  {\bibfnamefont {M.~G.}\ \bibnamefont {Vergniory}}, \bibinfo {author}
  {\bibfnamefont {Z.}~\bibnamefont {Wang}}, \bibinfo {author} {\bibfnamefont
  {C.}~\bibnamefont {Felser}}, \bibinfo {author} {\bibfnamefont {M.~I.}\
  \bibnamefont {Aroyo}}, \ and\ \bibinfo {author} {\bibfnamefont {B.~A.}\
  \bibnamefont {Bernevig}},\ }\href {http://dx.doi.org/10.1038/nature23268}
  {\bibfield  {journal} {\bibinfo  {journal} {Nature}\ }\textbf {\bibinfo
  {volume} {547}},\ \bibinfo {pages} {298} (\bibinfo {year}
  {2017})}\BibitemShut {NoStop}%
\bibitem [{\citenamefont {Slager}\ \emph {et~al.}(2015)\citenamefont {Slager},
  \citenamefont {Rademaker}, \citenamefont {Zaanen},\ and\ \citenamefont
  {Balents}}]{Slager2015}%
  \BibitemOpen
  \bibfield  {author} {\bibinfo {author} {\bibfnamefont {R.-J.}\ \bibnamefont
  {Slager}}, \bibinfo {author} {\bibfnamefont {L.}~\bibnamefont {Rademaker}},
  \bibinfo {author} {\bibfnamefont {J.}~\bibnamefont {Zaanen}}, \ and\ \bibinfo
  {author} {\bibfnamefont {L.}~\bibnamefont {Balents}},\ }\href {\doibase
  10.1103/PhysRevB.92.085126} {\bibfield  {journal} {\bibinfo  {journal}
  {Physical Review B}\ }\textbf {\bibinfo {volume} {92}},\ \bibinfo {pages}
  {085126} (\bibinfo {year} {2015})}\BibitemShut {NoStop}%
\bibitem [{\citenamefont {Hatsugai}(1993)}]{PhysRevLett.71.3697}%
  \BibitemOpen
  \bibfield  {author} {\bibinfo {author} {\bibfnamefont {Y.}~\bibnamefont
  {Hatsugai}},\ }\href {\doibase 10.1103/PhysRevLett.71.3697} {\bibfield
  {journal} {\bibinfo  {journal} {Phys. Rev. Lett.}\ }\textbf {\bibinfo
  {volume} {71}},\ \bibinfo {pages} {3697} (\bibinfo {year}
  {1993})}\BibitemShut {NoStop}%
\bibitem [{\citenamefont {Rhim}\ \emph {et~al.}(2018)\citenamefont {Rhim},
  \citenamefont {Bardarson},\ and\ \citenamefont {Slager}}]{Rhim2018}%
  \BibitemOpen
  \bibfield  {author} {\bibinfo {author} {\bibfnamefont {J.-W.}\ \bibnamefont
  {Rhim}}, \bibinfo {author} {\bibfnamefont {J.~H.}\ \bibnamefont {Bardarson}},
  \ and\ \bibinfo {author} {\bibfnamefont {R.-J.}\ \bibnamefont {Slager}},\
  }\href@noop {} {\bibfield  {journal} {\bibinfo  {journal} {Physical Review
  B}\ }\textbf {\bibinfo {volume} {97}},\ \bibinfo {pages} {115143} (\bibinfo
  {year} {2018})}\BibitemShut {NoStop}%
\bibitem [{\citenamefont {Slager}\ \emph {et~al.}(2016)\citenamefont {Slager},
  \citenamefont {Juri\ifmmode \check{c}\else \v{c}\fi{}i\ifmmode~\acute{c}\else
  \'{c}\fi{}}, \citenamefont {Lahtinen},\ and\ \citenamefont
  {Zaanen}}]{PhysRevB.93.245406}%
  \BibitemOpen
  \bibfield  {author} {\bibinfo {author} {\bibfnamefont {R.-J.}\ \bibnamefont
  {Slager}}, \bibinfo {author} {\bibfnamefont {V.}~\bibnamefont {Juri\ifmmode
  \check{c}\else \v{c}\fi{}i\ifmmode~\acute{c}\else \'{c}\fi{}}}, \bibinfo
  {author} {\bibfnamefont {V.}~\bibnamefont {Lahtinen}}, \ and\ \bibinfo
  {author} {\bibfnamefont {J.}~\bibnamefont {Zaanen}},\ }\href {\doibase
  10.1103/PhysRevB.93.245406} {\bibfield  {journal} {\bibinfo  {journal} {Phys.
  Rev. B}\ }\textbf {\bibinfo {volume} {93}},\ \bibinfo {pages} {245406}
  (\bibinfo {year} {2016})}\BibitemShut {NoStop}%
\bibitem [{\citenamefont {{Zirnstein}}\ \emph {et~al.}(2019)\citenamefont
  {{Zirnstein}}, \citenamefont {{Refael}},\ and\ \citenamefont
  {{Rosenow}}}]{nonherm_green}%
  \BibitemOpen
  \bibfield  {author} {\bibinfo {author} {\bibfnamefont {H.-G.}\ \bibnamefont
  {{Zirnstein}}}, \bibinfo {author} {\bibfnamefont {G.}~\bibnamefont
  {{Refael}}}, \ and\ \bibinfo {author} {\bibfnamefont {B.}~\bibnamefont
  {{Rosenow}}},\ }\href@noop {} {\bibfield  {journal} {\bibinfo  {journal}
  {arXiv preprint arXiv:1901.11241}\ } (\bibinfo {year} {2019})}\BibitemShut
  {NoStop}%
\bibitem [{\citenamefont {Essin}\ and\ \citenamefont
  {Gurarie}(2011)}]{essin2011bulk}%
  \BibitemOpen
  \bibfield  {author} {\bibinfo {author} {\bibfnamefont {A.~M.}\ \bibnamefont
  {Essin}}\ and\ \bibinfo {author} {\bibfnamefont {V.}~\bibnamefont
  {Gurarie}},\ }\href@noop {} {\bibfield  {journal} {\bibinfo  {journal}
  {Physical Review B}\ }\textbf {\bibinfo {volume} {84}},\ \bibinfo {pages}
  {125132} (\bibinfo {year} {2011})}\BibitemShut {NoStop}%
\bibitem [{\citenamefont {Bernevig}\ and\ \citenamefont
  {Hughes}(2013)}]{bernevig2013topological}%
  \BibitemOpen
  \bibfield  {author} {\bibinfo {author} {\bibfnamefont {B.~A.}\ \bibnamefont
  {Bernevig}}\ and\ \bibinfo {author} {\bibfnamefont {T.~L.}\ \bibnamefont
  {Hughes}},\ }\href@noop {} {\emph {\bibinfo {title} {Topological insulators
  and topological superconductors}}}\ (\bibinfo  {publisher} {Princeton
  university press},\ \bibinfo {year} {2013})\BibitemShut {NoStop}%
\bibitem [{\citenamefont {Roy}\ and\ \citenamefont
  {Harper}(2017)}]{PhysRevB.96.155118}%
  \BibitemOpen
  \bibfield  {author} {\bibinfo {author} {\bibfnamefont {R.}~\bibnamefont
  {Roy}}\ and\ \bibinfo {author} {\bibfnamefont {F.}~\bibnamefont {Harper}},\
  }\href {\doibase 10.1103/PhysRevB.96.155118} {\bibfield  {journal} {\bibinfo
  {journal} {Phys. Rev. B}\ }\textbf {\bibinfo {volume} {96}},\ \bibinfo
  {pages} {155118} (\bibinfo {year} {2017})}\BibitemShut {NoStop}%
\bibitem [{\citenamefont {Heeger}\ \emph {et~al.}(1988)\citenamefont {Heeger},
  \citenamefont {Kivelson}, \citenamefont {Schrieffer},\ and\ \citenamefont
  {Su}}]{RevModPhys.60.781}%
  \BibitemOpen
  \bibfield  {author} {\bibinfo {author} {\bibfnamefont {A.~J.}\ \bibnamefont
  {Heeger}}, \bibinfo {author} {\bibfnamefont {S.}~\bibnamefont {Kivelson}},
  \bibinfo {author} {\bibfnamefont {J.~R.}\ \bibnamefont {Schrieffer}}, \ and\
  \bibinfo {author} {\bibfnamefont {W.~P.}\ \bibnamefont {Su}},\ }\href
  {\doibase 10.1103/RevModPhys.60.781} {\bibfield  {journal} {\bibinfo
  {journal} {Rev. Mod. Phys.}\ }\textbf {\bibinfo {volume} {60}},\ \bibinfo
  {pages} {781} (\bibinfo {year} {1988})}\BibitemShut {NoStop}%
\bibitem [{\citenamefont {Hirsbrunner}\ \emph {et~al.}(2019)\citenamefont
  {Hirsbrunner}, \citenamefont {Philip},\ and\ \citenamefont
  {Gilbert}}]{hirsbrunner2019topology}%
  \BibitemOpen
  \bibfield  {author} {\bibinfo {author} {\bibfnamefont {M.~R.}\ \bibnamefont
  {Hirsbrunner}}, \bibinfo {author} {\bibfnamefont {T.~M.}\ \bibnamefont
  {Philip}}, \ and\ \bibinfo {author} {\bibfnamefont {M.~J.}\ \bibnamefont
  {Gilbert}},\ }\href@noop {} {\bibfield  {journal} {\bibinfo  {journal} {arXiv
  preprint arXiv:1901.09961}\ } (\bibinfo {year} {2019})}\BibitemShut {NoStop}%
\bibitem [{\citenamefont {Chen}\ and\ \citenamefont
  {Zhai}(2018)}]{chen2018hall}%
  \BibitemOpen
  \bibfield  {author} {\bibinfo {author} {\bibfnamefont {Y.}~\bibnamefont
  {Chen}}\ and\ \bibinfo {author} {\bibfnamefont {H.}~\bibnamefont {Zhai}},\
  }\href@noop {} {\bibfield  {journal} {\bibinfo  {journal} {Physical Review
  B}\ }\textbf {\bibinfo {volume} {98}},\ \bibinfo {pages} {245130} (\bibinfo
  {year} {2018})}\BibitemShut {NoStop}%
\bibitem [{\citenamefont {Porras}\ and\ \citenamefont
  {Fernandez-Lorenzo}(2018)}]{porras2018topological}%
  \BibitemOpen
  \bibfield  {author} {\bibinfo {author} {\bibfnamefont {D.}~\bibnamefont
  {Porras}}\ and\ \bibinfo {author} {\bibfnamefont {S.}~\bibnamefont
  {Fernandez-Lorenzo}},\ }\href@noop {} {\bibfield  {journal} {\bibinfo
  {journal} {arXiv preprint arXiv:1812.01348}\ } (\bibinfo {year}
  {2018})}\BibitemShut {NoStop}%
\bibitem [{\citenamefont {Wang}\ and\ \citenamefont
  {Zhang}(2012)}]{wang2012simplified}%
  \BibitemOpen
  \bibfield  {author} {\bibinfo {author} {\bibfnamefont {Z.}~\bibnamefont
  {Wang}}\ and\ \bibinfo {author} {\bibfnamefont {S.-C.}\ \bibnamefont
  {Zhang}},\ }\href@noop {} {\bibfield  {journal} {\bibinfo  {journal}
  {Physical Review X}\ }\textbf {\bibinfo {volume} {2}},\ \bibinfo {pages}
  {031008} (\bibinfo {year} {2012})}\BibitemShut {NoStop}%
\bibitem [{\citenamefont {Fukui}\ \emph {et~al.}(2008)\citenamefont {Fukui},
  \citenamefont {Fujiwara},\ and\ \citenamefont
  {Hatsugai}}]{doi:10.1143/JPSJ.77.123705}%
  \BibitemOpen
  \bibfield  {author} {\bibinfo {author} {\bibfnamefont {T.}~\bibnamefont
  {Fukui}}, \bibinfo {author} {\bibfnamefont {T.}~\bibnamefont {Fujiwara}}, \
  and\ \bibinfo {author} {\bibfnamefont {Y.}~\bibnamefont {Hatsugai}},\ }\href
  {\doibase 10.1143/JPSJ.77.123705} {\bibfield  {journal} {\bibinfo  {journal}
  {Journal of the Physical Society of Japan}\ }\textbf {\bibinfo {volume}
  {77}},\ \bibinfo {pages} {123705} (\bibinfo {year} {2008})},\ \Eprint
  {http://arxiv.org/abs/https://doi.org/10.1143/JPSJ.77.123705}
  {https://doi.org/10.1143/JPSJ.77.123705} \BibitemShut {NoStop}%
\bibitem [{\citenamefont {Kaufmann}\ \emph {et~al.}(2016)\citenamefont
  {Kaufmann}, \citenamefont {Li},\ and\ \citenamefont
  {Wehefritz-Kaufmann}}]{doi:10.1142/S0129055X1630003X}%
  \BibitemOpen
  \bibfield  {author} {\bibinfo {author} {\bibfnamefont {R.~M.}\ \bibnamefont
  {Kaufmann}}, \bibinfo {author} {\bibfnamefont {D.}~\bibnamefont {Li}}, \ and\
  \bibinfo {author} {\bibfnamefont {B.}~\bibnamefont {Wehefritz-Kaufmann}},\
  }\href {\doibase 10.1142/S0129055X1630003X} {\bibfield  {journal} {\bibinfo
  {journal} {Reviews in Mathematical Physics}\ }\textbf {\bibinfo {volume}
  {28}},\ \bibinfo {pages} {1630003} (\bibinfo {year} {2016})},\ \Eprint
  {http://arxiv.org/abs/https://doi.org/10.1142/S0129055X1630003X}
  {https://doi.org/10.1142/S0129055X1630003X} \BibitemShut {NoStop}%
\bibitem [{\citenamefont {Kitaev}(2009{\natexlab{b}})}]{kitaev2009periodic}%
  \BibitemOpen
  \bibfield  {author} {\bibinfo {author} {\bibfnamefont {A.}~\bibnamefont
  {Kitaev}},\ }in\ \href@noop {} {\emph {\bibinfo {booktitle} {AIP Conference
  Proceedings}}},\ Vol.\ \bibinfo {volume} {1134}\ (\bibinfo {organization}
  {AIP},\ \bibinfo {year} {2009})\ pp.\ \bibinfo {pages} {22--30}\BibitemShut
  {NoStop}%
\bibitem [{\citenamefont {Lieu}(2018)}]{lieu2018topological}%
  \BibitemOpen
  \bibfield  {author} {\bibinfo {author} {\bibfnamefont {S.}~\bibnamefont
  {Lieu}},\ }\href@noop {} {\bibfield  {journal} {\bibinfo  {journal} {Physical
  Review B}\ }\textbf {\bibinfo {volume} {97}},\ \bibinfo {pages} {045106}
  (\bibinfo {year} {2018})}\BibitemShut {NoStop}%
\end{thebibliography}%

\section{Supplemental Material: Topological Significance of Green's function Zeros} \label{GFchernappend}
\indent \indent We illustrate the topological significance of Green's function zeros by computing the Chern number. Expressed in terms of Green's functions, 
\begin{eqnarray}\label{chernnumber}
\mathcal{C}_{\alpha} = N_{\alpha} \, \epsilon^{\mu,\nu,\ldots}\int d^{d}k \, d\omega \ \text{Tr}\left[\G\partial_{\mu}\G^{-1}\G\partial_{\nu}\G^{-1}\ldots \right] \nonumber\\
\end{eqnarray}
\no where $\alpha$ identifies the Chern number, $N_{d}$ some quantized constant, $d$ the dimension, and $\mu,\nu = 0,1,\ldots d$. 
The Chern number, $\mathcal{C}_{\alpha}$, is robust under small variations $\delta \G$ which keep $\G^{-1}$ finite and is hence a topological invariant \cite{bernevig2013topological,wang2012simplified}. A band Chern number only changes by crossing a Green's function zero, corresponding to a pole in integrand of  Eq. \ref{chernnumber} (order of zero determines change in $\mathcal{C}_{\alpha}$). This can be generalized to any topological invariant \cite{doi:10.1143/JPSJ.77.123705,doi:10.1142/S0129055X1630003X}. Then, by bulk-boundary correspondence, when applicable, in-gap zeros of the edge localized Green's function must correspond to the bulk Green's function zeros, and by extension also track the band Chern numbers.\\
\indent We compute the topological invariant of a Hermitian system by counting the inter-band zero crossings, along with the order of the zero, $z_{m}$, and remembering that the total bulk invariant for all bands must sum to zero, $\sum\mathcal{C}_{i} = 0$ and $\mathcal{C}_{i+1} - \mathcal{C}_{i} = \sum_{m} z_{m}$. In the non-Hermitian case, we use the doubled Hamiltonian to compute the band topology and then define the net topological invariant of each single sub-block. Hence, even for open systems, we determine the band topological invariants. 
\section{Supplemental Material: In-Gap Bound State Topology}
\indent We elaborate on the central result in the main text. As discussed above, any definition of a band gap in the complex plane is inherently ambiguous. We circumvent this concern via the construction of a real gap defined by the doubled Hamiltonian formalism \cite{2018arXiv181210490Z,kawabata2018symmetry,kawabata2019topological}.
\subsection{Doubled from Single Band gap}\label{bandappend}
\indent By construction, the doubled gap center will be $E = 0$, with one set of eigenvalues (without loss of generality, $\NH$) above $E= 0$ and the other ($\NH^{\dagger}$) below. In this way, the topology of a single band encircling a point is also well defined \cite{2018arXiv181210490Z,nonherm_green}. In this work we focus on in-gap bound states between two bands and their behavior as bands become inseparable.\\
\indent Consider first the two bands in Fig.~\ref{figure1}a. The topology of either band is defined with respect to crossing the line $E(x) = 0+ix$, and the doubled Hamiltonian will consist of four bands, two above ($\NH$) and two ($\NH^{\dagger}$) below the gap. Their positions on the real axis is given by the magnitude of their energies. The topology of each band is defined with respect to its counterpart below $E = 0$ \cite{2018arXiv181210490Z}.\\
\indent Next, we look at the inseparable \cite{Shen2018} bands in Fig.~\ref{figure1}b. The topology of the \textit{single band} need not be trivial \cite{nonherm_green}. Here the doubled system consists of two bands one above $E = 0$ and the other below. This occurs in our example, the non-Hermitian Chern insulator, for intermediate non-Hermiticity. In the doubled system we see clear topological edge modes corresponding to a non-trivial point gap invariant.\\
\indent Finally, we turn to the bands in Fig.~\ref{figure1}c. The topology of either band is defined with respect to the point gap it encircles and the band gap between the two bands. The doubled gap will correspond to the magnitude of the band energies, with two bands above $E = 0$ and two below. To detect a topological invariant with respect to the band gap, we consider the Green's function zeros at energies between the two bands. By contrast, to detect the point gap we consider energies across $E = 0$. \\
\subsection{Doubled - Single Green's Function Correspondence} \label{GFcorrappendix}
\indent We define in-gap bound states with respect to the doubled gap and connect them with single system bound states. In particular, we take advantage of the doubled Hamiltonian Hermiticity to define a doubled Green's function whose zeros have topological significance, and map them to corresponding zeros in the individual non-Hermitian systems.\\
\indent We begin with Eq. \eqref{doubledgreen},
\begin{align}
\GG (\omega) =	
\frac{ \GG_{0}(\omega}
{1 - \hat{\mathbb{V}}\, \GG_{0}(\omega)}, \ \ \text{with} \ \ 
 \GG_{0}(\omega) =	
\frac{1}{\omega - \HH_0}.\nonumber
\end{align}
%\noindent The Hermiticity of $\GG$ guarantees topological boundary modes correspond to zeros of $\GG_{0}$
\noindent By the Hermiticity of $\GG$, topological boundary modes correspond to zeros of $\GG_{0}$. Then, defining \textit{undoubled} (non-Hermitian) Green's functions for $\omega\in\mathbb{R}$ and $\theta\in[0,2\pi]$, we relate the zeros of $\GG_{0}$ to those of $\G_{0},\G_{0}^{\dagger}$,
\begin{eqnarray} \label{singleGFs}
\G_{0}(\omega,\theta)  \equiv (\omega e^{i\theta} - \NH_{0})^{-1}, \ \ \G^\dag_{0}(\omega,\text{-}\theta) \equiv(\omega e^{\text{-}i\theta} -\NH_{0}^{\dagger})^{-1}.\nonumber
\end{eqnarray}
\indent Note that $\GG_{0}$ is only defined for $\omega \in\mathbb{R}$, but the zeros of $\G_{0},\G_{0}^{\dagger}$ are, in general, complex. However, using that $\omega^{2} = (\omega e^{i\theta})(\omega e^{\text{-}i\theta})$ for any $\theta\in[0,2\pi]$, we attempt to factor out single Green's functions. We will always work at distance $\delta >0$ from any singlular points and then take $\delta\rightarrow 0$. This will not always be possible, as in the case of inseparable bands (Fig.~\ref{figure1}b), where projection to an edge may render $\G_{0}$  singular everywhere in-gap. In such cases, this factorization does not hold, see discussion below. Returning to our factorization,
\begin{widetext}
\begin{eqnarray}
\GG_{0}(\omega) &=& \left[ \omega\left(\begin{pmatrix}
1 & 0 \\ 0 & 1
\end{pmatrix} -\begin{pmatrix}
0 & e^{i\theta} \\ e^{\text{-}i\theta} & 0
\end{pmatrix} + \begin{pmatrix}
0 & e^{i\theta} \\ e^{\text{-}i\theta} & 0
\end{pmatrix} \right) -\begin{pmatrix}
0 & \NH_{0} \\ \NH^{\dagger}_{0} & 0 
\end{pmatrix}\right]^{-1} = \left[ \omega\begin{pmatrix}
1 & \text{-}e^{i\theta} \\ \text{-}e^{\text{-}i\theta} & 1
\end{pmatrix}  +\begin{pmatrix}
0 & e^{i\theta}\omega-\NH_{0} \\ e^{\text{-}i\theta}\omega-\NH^{\dagger}_{0} & 0 
\end{pmatrix}\right]^{-1}\nonumber\\
&=& \begin{pmatrix}
0 & e^{i\theta}\omega-\NH_{0} \\ e^{\text{-}i\theta}\omega-\NH^{\dagger}_{0} & 0 
\end{pmatrix}^{-1}\left[ \omega\begin{pmatrix}
0 & e^{i\theta}\omega-\NH_{0} \\ e^{\text{-}i\theta}\omega-\NH^{\dagger}_{0} & 0 
\end{pmatrix}^{-1}\begin{pmatrix}
1 & \text{-}e^{i\theta} \\ \text{-}e^{\text{-}i\theta} & 1
\end{pmatrix}  + 1\right]^{-1}.
\end{eqnarray}
Taking advantage of inverse properties of block off-diagonal matrices:
\begin{eqnarray}
\begin{pmatrix}
0 & A \\ B & 0 
\end{pmatrix}^{-1} = \begin{pmatrix}
0 & B^{-1}\\ A^{-1} & 0 
\end{pmatrix},
\end{eqnarray}
we obtain
\begin{align}\label{factoredG}
\GG_{0}(\omega) = \begin{pmatrix}
0 & (e^{\text{-}i\theta}\omega-\NH^{\dagger}_{0})^{-1}\\ (e^{i\theta}\omega-\NH_{0})^{-1}& 0 
\end{pmatrix}\left[ \omega\begin{pmatrix}
0 & (e^{\text{-}i\theta}\omega-\NH^{\dagger}_{0})^{-1}\\ (e^{i\theta}\omega-\NH_{0})^{-1}& 0 
\end{pmatrix}\begin{pmatrix}
1 & \text{-}e^{i\theta} \\ \text{-}e^{\text{-}i\theta} & 1
\end{pmatrix}  + 1\right]^{-1} .
\end{align}
Substituting the previously defined single Green's functions in Eq. \eqref{singleGFs} into Eq. \eqref{factoredG} yields 
\begin{align}\label{correspondence}
\GG_{0}(\omega) = \begin{pmatrix}
0 & \G_{0}^{\dagger}(\omega,\theta) \\ \G_{0}(\omega,\text{-}\theta) & 0
\end{pmatrix}
\left[1 +  \omega \begin{pmatrix}
0 & \G_{0}^{\dagger}(\omega,\text{-}\theta) \\ \G_{0}(\omega,\theta) & 0
\end{pmatrix} \begin{pmatrix}
1 & \text{-}e^{i\theta} \\ \text{-}e^{\text{-}i\theta} & 1
\end{pmatrix}  \right]^{-1}.
\end{align}
In Eq. \eqref{correspondence} above, we write $\G_{0}(\omega,\theta)$ to indicate the domain change into the complex plane, with $\omega \in \mathbb{R}$, s$\theta\in[0,2\pi]$. We are free to take any choice $\theta$ in Eq. \eqref{correspondence}. So, for each $\omega_{*} \in \mathbb{R}$ defined by  $\G_{0}(\omega_{*},\theta_{*}) = \G_{0}^{\dagger}(\omega_{*},\text{-}\theta_{*}) = 0$, we take $\theta(\omega_{*}) =\theta_{*}$, arbitrarily choosing the smallest $\theta_{*}$ in case of degeneracy. Otherwise, we arbitrarily choose $\theta = 0$. This defines a path in $\omega$ for $\theta$,
\begin{eqnarray}\label{pathinC}
 \theta:\mathbb{R}\rightarrow[0,2\pi]
  \quad  \Bigg{\vert} \quad
   \theta(\omega) = 
  \begin{cases} 
\theta_{*}, & \text{if, } \exists\theta_{*}\in[0,2\pi],\ \text{such that } \G_{0}(\omega,\theta_{*})  = 0\\
0, & \text{otherwise }
\end{cases},\quad  \omega\in\mathbb{R}.
\end{eqnarray} 
Therefore,  as $\omega \rightarrow\omega_{*}$, $\GG_{0} (\omega)\rightarrow 0$. We remark that if for any $\omega_{*}\in\mathbb{R}$, there exists a $\theta_{*}\in [0,2\pi]$ such that $\G_{0}(\omega,\theta) = 0$, then $\GG_{0}(\omega_{*})$ has a zero eigenvalue. Zeros are a special factorization in this sense. By contrast, if $\theta = \theta_{nz}$ where $\G_{0}(\omega_{*},\theta_{nz})\neq 0$, this guarantees the first term in Eq. \eqref{correspondence} is not singular, and we can recombine $\GG_{0}$ back into the form of Eq. \eqref{doubledgreen}, which by the $\theta_{*}$ factorization above must have a zero eigenvalue. Note, in the case of degenerate zeros, either factorization guarantees a zero of $\GG_{0}$. In this way, we construct a correspondence between the zeros of the single Green's function and the doubled Green's function. Here all topological features are defined by the doubled Green's function zeros and through this correspondence are inherited by the single Green's function zeros.\\
\indent We stress the importance of the band gap. At all times our procedure requires a gap to be well defined in the complex plane. This is particularly important in the case of a point gap, Fig.~\ref{figure1}b, where the doubled Green's function gap is defined between two bands ($\NH$ and $\NH^{\dagger}$). A single band may carry a non-trivial topological invariant \cite{2018arXiv181210490Z}, but the edge projection may collapse the point gap. In such cases the single band Green's function is trivially zero in the absence of a gap. Instead, the doubled Green's function has ``in-gap solutions" corresponding to region between the $\NH$, $\NH^{\dagger}$ bands. The single band system does not have a corresponding topological bound state and is singular everywhere. This is precisely case III of Table \ref{tabulate}, the trivial skin effect. \end{widetext}

\section{Supplemental Material: Generalized Bulk-Boundary Correspondence}\label{generalizedappend}
\indent The primary content of this work is presented in Table \ref{tabulate}. The formalism developed in this work generalizes the notion of bulk-boundary correspondence to generic non-Hermitian systems. This generalization has some technical caveats discussed here.\\
\indent First, the doubled Green's function has a well defined real gap between a single band and its complex conjugate, see \ref{bandappend}. Bound states of the doubled system are well defined, but this does not, in general, translate into a bulk-boundary correspondence for a single topological band. In that sense, non-Hermitian systems may carry bulk topological invariants for which bulk-boundary correspondence is broken. However, in the presence of a well defined band gap, Fig.~\ref{figure1}a and c, our work restores a generalized form of bulk-boundary correspondence, cases I and II. We discuss point gaps in section \ref{appendPoint}.\\
\indent While there exists a well defined band gap, the single Green's function is non-singular almost everywhere and inverses are well defined. By Eq. \eqref{baredGF} we can always relate the zeros of the single and doubled Green's functions. The in-gap zeros of the single Green's function correspond to bound states of the system, but non-Hermitian Green's function zeros are not universal due to the localization of bulk modes to an open boundary, the non-Hermitian skin effect \cite{PhysRevLett.121.136802,lee2018anatomy}, see \ref{appendPoint}. However, through this correspondence, we can identify edge localized modes that correspond to in-gap zeros of the doubled Green's function. In previous literature, when the single Green's function zeros are constrained to the real energy axis \cite{Kunst2018b,Kawabata2018}, the system was said to preserve bulk-boundary correspondence. In particular, the edge mode is present on both edges of the system, and the zero-energy mode is protected by the doubled gap from non-universal zeros \cite{PhysRevLett.121.136802}. However, our formalism distinguishes topological complex energy edge states from trivial ones. These may not be separated in energy substantially from the non-universal bound states, but are nonetheless topological. We distinguish these modes from the non-Hermitian skin effect as the anomalous skin effect, case II in Table \ref{tabulate}. Therefore in the presence of a band gap, non-trivial band topology is protected, and we can always define an in-gap edge mode reflecting the bulk topological degeneracy. In this sense, we generalize the notion of bulk-boundary correspondence to non-Hermitian systems. Symmetry constraints can restore traditional bulk-boundary correspondence, but the topological property exists beyond the symmetries of the model. This is intimately tied to the segmentation of the complex plane by both a line and a band gap as illustrated in Fig.~\ref{figure1}. This gives a direct geometric interpretation to our exhaustive classification of bulk boundary correspondences in non-Hermitian systems.

\subsection{Point Gap and Beyond Bulk Boundary Correspondence}\label{appendPoint}
\indent We discuss case III in Table \ref{tabulate}. As mentioned above, point gap topology ascribes a net non-trivial bulk topology, e.g. a net winding of the band. In the absence of a line or band gap protecting the bulk topology, non-Hermitian bands can still exhibit a point gap topological invariant associated with the structure of the band on the complex plane. Non-Hermitian energy bands are closed loops which generically enclose non-zero area in the complex plane, e.g. $E = t e^{ik}$ for a 1-D one way hopping model, $\NH = \sum_{x} t\hat{c}^{\dagger}_{x+1}\hat{c}_{x}$. However, unlike line or band gaps, point gaps do not generically protect the bulk topological invariant in the presence of an edge. In particular, an edge projects the bulk Hamiltonian. Heuristically, closed loops may be cut and projected into lines or points, making them trivially contractible to a point. Such an edge non-locally trivializes the point gap invariant of the entire bulk. Hence, all bulk information must be localized at that edge, resulting in a localization of all bulk modes at the edge. For example, if the band had net winding, broken by a right and left edge, all bulk modes can only move from one edge to the other, localizing them at the terminating edge. In fact, for the the one-way hopping model, we see precisely this phenomenon, manifested as an N-th order \textit{exceptional point} -- the N dimensional system has rank 1, meaning every state is localized to one site \cite{Kunst2018a,PhysRevE.59.6433}. This is what we refer to as the trivial non-Hermitian skin effect, trivial because localized modes do not reflect the bulk topological invariant. \\
\indent For example, the non-Hermitian Chern insulator when projected to the edge, $x = 0$, is equivalent to the non-Hermitian SSH model. The bulk point gap invariant is equivalent for $h_{z} \neq 0$ or $h_{y} \neq 0$ (see Supplemental Material \ref{chernsymappend}), but the edge spectrum for $h_{y}\neq 0$ is pseudo-Hermitian, which implies either purely real or purely imaginary Eigenvalues. Hence, the bands are flattened to lines in the complex plane and point gap topology disappears. By contrast, if $h_{z} \neq 0$, the projected Hamiltonian does not flatten the bands in the complex plane. They form a point gap, and the topology is not trivialized. Thus, for $h_{y}\neq0$ we see the trivial non-Hermitian skin effect. By contrast, for $h_{z}\neq 0$, we see no localization of bulk edge modes as the bulk point gap topology is not trivialized by the edge. In fact, we see topological (anomalous skin) edge modes persisting in this regime despite the gapless bulk \cite{Kawabata2018}, corresponding to the point gap topology of the bulk, Supplemental Material \ref{chernsymappend}.

\section{Supplemental Material: Non-Hermitian Chern Insulator}\label{chernappend}
We elaborate some computational details. In particular, we obtain the restricted single Green's function analytically by the same technique used in \cite{Slager2015}. The appearance and disappearance of the restricted Green's function zeros are then used to determine the Chern number and other properties of the non-Hermitian Chern insulator.
\subsection{Green's Function Poles}\label{chernpolesappend}
In this section show the analytic results on the non-Hermitian Chern insulator Green's function zeros that aided numerical results. We consider the Hamiltonian of form:

\begin{align}
\mathcal H = \vv \xi  \, \vv \sigma 
+ \vv \eta \, \vv \sigma 
+ i \, \vec h \,  \vv \sigma, 
\end{align}

\no
with

\begin{align}
\vv \xi & = (\cos k_x, -\sin k_x, 0), 
\\
\vv \eta & = (\cos k_y - m, 0 , \sin k_y),
\\
\vec h & = (h_x, h_y, h_z), 
\end{align}
Note we choose this notation in this section to maximize carry-over between the non-Hermitian Chern insulator and the SSH model, but combine the Hermitian terms $\vv \xi+ \vv \eta\rightarrow\vv \xi$ in the remainder of the text.\\
\indent We now construct the Green's function
\begin{align}
\G (\omega, \vec k)   
=  \frac{1}{\omega - \mathcal H}
=  \frac{\omega + \mathcal H}{\omega^2 - \mathcal H^2}.
\end{align}
\no We explicitly calculate the denominator
\begin{align}
\mathcal H ^2 & = (\vv \xi  \, \vv \sigma 
+ \vv \eta \, \vv \sigma 
+ i \, \vec h \,  \vv \sigma)^2  
\nonumber
\\
%& = (\vv \xi +  \vv \eta )^2  - \vec h^2 
%+ 2 (\xi_x \eta_x + \xi_y \eta_y)
%+ 2i (\xi_x h_x + \xi_y h_y)
%\nonumber
%\\
& = \Omega^2 + 2 \cos{k_x} (\eta_x + i h_x)
 - 2 \sin k_x (\eta_y + i h_y),  
\end{align}
\no where 
\begin{align}
\Omega^2 = 1 + \eta^2 - h^2 + 2 i \, \vv \eta \, \vec h. 
\end{align}
\no We substitute $t \equiv e^{i k_x}$, such that $\cos k_x =  (t+ 1/t)/2$ and $\sin k_x =  (t - 1/t)/2 i$. Thus,
\begin{align}
\label{Hsqr}
\mathcal{H}^{2}=\Omega^2 + (\alpha + i \beta ) t + (\alpha^* + i \, \beta^*) t^{-1}, 
\end{align}
\no where we have introduced the following quantities:
\begin{align}
\alpha & = \eta_x + i \eta_y,
\\
\beta & = h_x + i h_y,
\end{align}
\indent We Fourier transform and restrict $\vec r_{\bot} = 0$,
\begin{align}
\G (\omega, k_y, \vec r_{\bot} = 0)   
=  \int \limits_{-\pi}^{+\pi} \frac{d k_x}{2 \pi}  \frac{\omega + \mathcal H}{\omega^2 - \mathcal H^2},
\end{align}
\no
Taking $t = e^{i k_x}$, we obtain the following countour integral around the unit circle:
\begin{align}
\G (\omega, k_y, \vec r_{\bot} = 0)   
=  \oint \limits_{\mathcal{S}} \frac{d t}{2 \pi i }  \frac{\omega + \mathcal H}{t (\omega^2 - \mathcal H^2)},
\end{align}
\no Using Eq. \eqref{Hsqr}, we can write
\begin{align}
\G (\omega, k_y, & \vec r_{\bot} = 0)  = 
\nonumber
\\ 
& = 
- \frac{1}{(\alpha + i \beta )}
\oint \frac{d t}{2 \pi i }  \frac{\omega + \mathcal H}
{  t^2  - 2 f(\omega) t +c}
\nonumber
\end{align}
\no where
\begin{align}
f(\omega)  = \frac{\omega^2 - \Omega^2 }
{2 (\alpha + i \beta )},
\ \ \  \ \ \ 
c  = \frac{\alpha^* + i \beta^* }{\alpha + i \beta }, 
\end{align}
\no Considering the integrand,
\begin{align}
Z(t, \omega) =  t^2  - 2 f(\omega) t +c = 0,
\end{align}
\no We find the roots,
\begin{align}
t_{1,2} = f (\omega) \pm \gamma (\omega),
 \ \ \  \ \  \text{with} \ \  
 \gamma(\omega) = \sqrt{f^2(\omega) - c}.
 \nonumber
\end{align}
To calculate the Green's function we only have to evaluate the following three contour integrals:
\begin{align}
I_0  & = \oint\limits_{\mathcal{S}} \frac{dt }{2 \pi i} \, 
\frac{1}{Z(t,\omega)},
\\
I_1  & = \oint\limits_{\mathcal{S}} \frac{dt }{2 \pi i} \, 
\frac{t}{Z(t,\omega)},
\\
I_2  & = \oint\limits_{\mathcal{S}} \frac{dt }{2 \pi i} 
\frac{1}{t \, Z(t,\omega)},
\end{align}
We rewrite the Green's function,
\begin{align}
\G (\omega, k_y, & \vec r_{\bot} = 0)  =  
- \frac{1}{(\alpha + i \beta )}
\G' (\omega, k_y) ,
\nonumber
\end{align}
\no with 
\begin{align}
 \G' (\omega, k_y)  
 & =  
I_0 (\omega \, \sigma_0
+
\vv \eta \, \sigma 
+ i \, \vec h \, \sigma
) 
\nonumber
\\
& + 
\frac{I_1} {2} (\sigma_x + i \sigma_y) 
+
\frac{I_2} {2} (\sigma_x - i \sigma_y) ,
\label{G(w,k)}
\end{align}
\no We use Cauchy's theorem to evaluate the contour integrals, which have in general up to 2 (3 for $I_2$) poles. There are several distinct cases:
\subsubsection{Both poles outside} 
\indent In the case when both poles $t_{1,2} = f(\omega) \pm \gamma (\omega)$ are outside the unit circle, there are no residues inside the unit circle and we have 
\begin{align}
I_0 = 0, 
\ \ \
I_1 = 0. 
\end{align}
\no However, the remaining integral is non-vanishing because it has a single pole at $t= 0$. 
\begin{align}
I_2   = \oint \frac{dt }{2 \pi i} 
\frac{1}{t \, Z(t)} = \res_{t = 0} \frac{1}{t \, Z(t)} = \frac{1}{c}. 
\end{align}
Thus, according to \eqref{G(w,k)},  in this case the Green's function has form
\begin{align}
\G'(\omega, k_y) = \frac{I_2}{2} (\sigma_x - i \sigma_y) = \frac{1}{c} 
\begin{pmatrix}
0 & 0
\\
1 & 0
\end{pmatrix},
\nonumber
\end{align}
\no and its eigenvalues are zero everywhere in the gap, i.e. the bands are touching.
\subsubsection{Both poles inside}
\indent In case when we have both poles inside the unit circle $|t|<1$ we have $I_0$ vanishing as both residues exactly cancel each other:
\begin{align}
I_0 & = \oint \frac{dt }{2 \pi i} \, 
\frac{1}{Z(t,\omega)} = \res_{t= t_1} \frac{1}{Z(t,\omega)} + \res_{t= t_2} \frac{1}{Z(t,\omega)} = 
\nonumber
\\
& = \frac{1}{t_1-t_2}+\frac{1}{t_2-t_1} = 0. 
\end{align}
\no The $I_1$ integral is however finite:
\begin{align}
I_1 & = \oint \frac{dt }{2 \pi i} \, 
\frac{t}{Z(t,\omega)} = \res_{t= t_1} \frac{t}{Z(t,\omega)} + \res_{t= t_2} \frac{t}{Z(t,\omega)} = 
\nonumber
\\
& = \frac{t_1}{t_1-t_2}+\frac{t_2}{t_2-t_1} = \frac{t_1- t_2}{t_1- t_2} = 1. 
\end{align}
The last integral is however zero:
\begin{align}
I_2 & = \oint \frac{dt }{2 \pi i} \, 
\frac{1}{t Z(t,\omega)} 
\nonumber
\\
& = \res_{t= t_1} \frac{1}{t Z(t,\omega)} + \res_{t= t_2} \frac{1}{t Z(t,\omega)}  +  \res_{t= 0} \frac{1}{t Z(t,\omega)}
\nonumber
\\
& = \frac{1}{t_1(t_1-t_2)}+\frac{1}{t_2 (t_2-t_1)} + \frac{1}{t_1 t_2} 
\nonumber
\\
& = 
\frac{t_2 - t_1 + t_1 - t_2}{t_1 t_2 (t_1 - t_2)} = 0. 
\end{align}
\no Therefore, the Green's function reads:
\begin{align}
\G'(\omega, k_y) = \frac{I_1}{2} (\sigma_x + i \sigma_y) =  
\begin{pmatrix}
0 & 1
\\
0 & 0
\end{pmatrix},
\nonumber
\end{align}
\no and its eigenvalues are also zero in this case, corresponding to the gapless case again.
\subsubsection{Both poles on the circle}
\indent In case when both poles are on circle, poles cancel each other again, and the system is gapless.
\subsubsection{First only inside}
\indent We now consider that only the first pole $t_1 = f(\omega) + \gamma(\omega)$ to be inside the unit circle. In this case have:
\begin{align}
I_0 & = \oint \frac{dt }{2 \pi i} \, 
\frac{1}{Z(t,\omega)} = \res_{t= t_1} \frac{1}{Z(t,\omega)}  
= \frac{1}{2 \, \gamma (\omega) }  ,
\nonumber
\end{align}
\begin{align}
I_1 & = \oint \frac{dt }{2 \pi i} \, 
\frac{t}{Z(t,\omega)} = \res_{t= t_1} \frac{t}{Z(t,\omega)}  
= \frac{f(\omega) + \gamma( \omega)}{2 \, \gamma (\omega) }
\nonumber
\\
& = \frac{t_1}{ 2 \gamma (\omega) } ,
\nonumber
\end{align}
\begin{align}
I_2 & = \oint \frac{dt }{2 \pi i} \, 
\frac{1}{t Z(t,\omega)} = \res_{t= t_1} \frac{t}{Z(t,\omega)}  
= \frac{1}{2 \, \gamma (\omega) \left (f(\omega) - \gamma( \omega) \right) }
\nonumber
\\
& = \frac{1}{2 \gamma (\omega) t_2}. 
\nonumber
\end{align}
Thus the Green's function reads:
\begin{align}
 \G' (\omega, k_y)  
 & =  
I_0 (\omega \, \sigma_0
+
\vv \eta \, \sigma 
+ i \, \vec h \, \sigma
) 
\nonumber
\\
& + 
\frac{I_1} {2} (\sigma_x + i \sigma_y) 
+
\frac{I_2} {2} (\sigma_x - i \sigma_y) ,
\nonumber
\\
&  = 
\frac{1}{2 \, \gamma (\omega) } 
\begin{pmatrix}
\omega + \delta & \alpha^* +  i \beta^{*} \\
\alpha +  i \beta & \omega - \delta
\end{pmatrix}
\nonumber
\\
 & + 
 \frac{t_1}{2 \, \gamma (\omega) }
\begin{pmatrix}
0 & 1 \\
0 & 0
\end{pmatrix}
 + 
 \frac{1}{2 \, \gamma (\omega) t_2}
\begin{pmatrix}
0 & 0 \\
1 & 0
\end{pmatrix}
\nonumber
\\
&  = 
\frac{1}{2 \, \gamma (\omega) } 
\begin{pmatrix}
\omega + \delta & \alpha^* +  i \beta^{*} + t_1\\
\alpha +  i \beta + \frac{1}{t_2} & \omega - \delta
\end{pmatrix} .
\nonumber
\end{align}
\noindent
with $\delta = \eta_z + i h_z$.
\subsubsection{Second only inside}
\indent We consider now that only  the second pole $t_1 = f(\omega) - \gamma(\omega)$ is placed inside the unit circle. In this case have:
\begin{align}
I_0 & = \oint \frac{dt }{2 \pi i} \, 
\frac{1}{Z(t,\omega)} = \res_{t= t_2} \frac{1}{Z(t,\omega)}  
=  - \frac{1}{2 \, \gamma (\omega) }. 
\nonumber
\end{align}
\begin{align}
I_1 & = \oint \frac{dt }{2 \pi i} \, 
\frac{t}{Z(t,\omega)} = \res_{t= t_2} \frac{t}{Z(t,\omega)}  
= - \frac{f(\omega) - \gamma( \omega)}{2 \, \gamma (\omega) }
\nonumber
\\
& = 
- \frac{t_2}{2 \gamma(\omega)} .
\nonumber
\end{align}
\begin{align}
I_2 & = \oint \frac{dt }{2 \pi i} \, 
\frac{1}{t Z(t,\omega)}  = \res_{t= t_2} \frac{t}{Z(t,\omega)}  
\nonumber
\\
& = -  \frac{1}{2 \, \gamma (\omega) \left (f(\omega) - \gamma( \omega) \right) } 
\nonumber
=
-\frac{1}{2 \gamma (\omega) t_1 }. 
\nonumber
\end{align}
Thus the Green's function in this case reads:
\begin{align}
 \G' (\omega, k_y)  
 & =  
I_0 (\omega \, \sigma_0
+
\vv \eta \, \sigma 
+ i \, \vec h \, \sigma
) 
\nonumber
\\
& + 
\frac{I_1} {2} (\sigma_x + i \sigma_y) 
+
\frac{I_2} {2} (\sigma_x - i \sigma_y) ,
\nonumber
\\
&  = 
 - \frac{1}{2 \, \gamma (\omega) } 
\begin{pmatrix}
\omega  + \delta & \alpha^* +  i \beta^{*} \\
\alpha +  i \beta & \omega - \delta
\end{pmatrix}
\nonumber
\\
 &  
 -  \frac{t_2}{2 \, \gamma (\omega) }
\begin{pmatrix}
0 & 1 \\
0 & 0
\end{pmatrix}
-  
 \frac{1}{2 \, \gamma (\omega) t_1}
\begin{pmatrix}
0 & 0 \\
1 & 0
\end{pmatrix}
\nonumber
\\
&  = 
 - \frac{1}{2 \, \gamma (\omega) } 
\begin{pmatrix}
\omega + \delta & \alpha^* +  i \beta^{*} + t_2\\
\alpha +  i \beta + \frac{1}{t_1} & \omega - \delta
\end{pmatrix} .
\nonumber
\end{align}
We see that the structure of the Green's function is similar to the previous case. 
\subsubsection{Eigenvalues of the Green's functions}
\indent To properly account for which pole is within the unit circle in numerics, we introduce quantity
\begin{align}
s(\omega) = \text{sign} \left( 1 - |t_{1} (\omega) | \right),
\end{align}
\no then if one of the poles 
\begin{align}
T_{1,2} = f (\omega) \pm s(\omega) \gamma (\omega),
\end{align}
\no is inside, we automatically guarantee it is $T_1$ which is always inside. 
Then the Green's function is given by 
\begin{align}
\label{Chern_GF}
 \G' (\omega, k_y)  
  = 
\frac{s (\omega) }{2 \, \gamma (\omega) } 
\begin{pmatrix}
\omega  + \delta& \alpha^* +  i \beta^{*} + T_1 (\omega)\\
\alpha +  i \beta +\frac{1}{T_2 (\omega)} & \omega - \delta
\end{pmatrix} .
\end{align}
The zero eigenvalues of this Green's function is given by $\text{det} \G(\omega ,k_y) = 0$, and yields 
\begin{align}
\omega^2  - \delta^2& = (\alpha + i \beta )(\alpha^* +  i \beta^{*})  
+ 
(\alpha + i \beta ) T_1 (\omega) 
\nonumber
\\
& + \frac{(\alpha^* +  i \beta^{*})}{T_2(\omega)}
+ 
\frac{T_{1} (\omega)}{T_{2} (\omega)} .
\label{zeros_GF}
\end{align}
\no We simplify this expression by noticing that
\begin{align}
\omega^2- \Omega^2 - (\alpha + i \beta ) T - (\alpha^* + i \, \beta^*) \frac{1}{T} = 0,  	
\end{align}
\no Therefore,
\begin{align}
\frac{(\alpha^* +  i \beta^{*})}{T_2(\omega)} = 
\omega^2 - \Omega^2 - (\alpha + i \beta ) T_2 (\omega) .
\end{align}
Then, taking now into account that 
\begin{align}
\frac{T_1 (\omega) }{T_2 (\omega)}
&  = 
\frac{f (\omega) + s(\omega) \gamma (\omega)}{f (\omega) - s(\omega) \gamma (\omega)} 
\nonumber	
\\
& = 
\nonumber
\frac{1}{c} \left[f (\omega) + s(\omega) \gamma (\omega) \right]^2 ,
\end{align}
\no and 
\begin{align}
T_1 (\omega) - T_2 (\omega)  = 2 s(\omega) \gamma (\omega) ,	
\end{align}
\no we have
\begin{align}
 \frac{1}{c} \left[f (\omega) + s(\omega) \gamma (\omega) \right]^2 +
2 (\alpha + i \beta ) s(\omega) \gamma(\omega)
\nonumber
\\
 = 
 \Omega^2 -  \delta^2 - |\alpha|^2 + |\beta|^2. 
\end{align}
\no By expanding and squaring this equation [to eliminate sign function $s (\omega)$], we get a polynomial (6th order) equation in $\omega$, which can be solved numerically. In fact, the formalism generically provides a dispersion relation for the edge mode.\\
\indent In the present case the dispersion is more simply computed by considering the Chern Hamiltonian as set of linked SSH Hamiltonians via projection onto the $\hat{y}$-axis,
\begin{eqnarray}
	H_{p} &=& \sum_{x} (\hat{c}_{k_{y},x}^{\dagger}\hat{c}_{k_{y},x+1}\sigma_{+} + H.c.) \nonumber\\
	&+& [(m-\cos(k_{y}),0,\sin(k_{y}))+\vv h]\cdot\vv \sigma \ \  \hat{c}_{k_{y},x}^{\dagger}\hat{c}_{k_{y},x}.Finally\nonumber\\
\end{eqnarray} 
\no For $h_{x},h_{y} = 0$, the existence of edge modes clearly corresponds to whether or not there exists a $k_{y}$ such that $m-1 -\cos k_{y} = 0$, i.e. $0<m<2$. And the dispersion relation is just the diagonal component of the projected Hamiltonian,
\begin{eqnarray}
\pm [\sin k_{y} + h_{z}].
\end{eqnarray} 
\no Finally, for $h_{y}\neq 0$, the condition for edge modes is simply $m-h_{y} -1 -\cos k_{y} = 0$.

\subsection{Symmetries}\label{chernsymappend}
Here we discuss the relevant symmetries of the non-Hermitian Chern insulator as parameterized by Eq. \eqref{Hamiltonian} of the main text for different non-Hermitian perturbations, $\vv h$, and compute the expected topological invariant of each set of symmetries in the presence of a point gap (see Fig.~\ref{figure1}b). Since we are interested in the bulk topology, we consider relevant symmetries. Inversion along $k_{y},k_{x}$ are symmetries of the system if respected by the non-Hermiticity. We use the conventions of \cite{2018arXiv181210490Z} to define symmetry actions on the Hamiltonian (chiral, $P$, pseudo-Hermiticity, $Q$, Transposition symmetry, $C$, Conjugation symmetry, $K$). Specifically,
\begin{eqnarray}
\NH(\vv k) &=& -\mathcal{P}\NH(a_{P}\vv k)\mathcal{P}^{-1},\mathcal{P}\mathcal{P}^{*} = \mathbb{I}, \ \ a_{P}=\pm1\nonumber\\
\NH(\vv k) &=& \mathcal{Q}\NH^{\dagger}(a_{Q}\vv k)\mathcal{Q}^{-1},\mathcal{Q}\mathcal{Q}^{*} = \mathbb{I}, \ \ a_{Q} = \pm1\nonumber\\
\NH(\vv k) &=& \epsilon_{C}\mathcal{C}\NH^{T}(a_{C}\vv k)\mathcal{C}^{-1},\mathcal{C}\mathcal{C}^{*} = \eta_{C}\mathbb{I}, \ \ \eta_{C},a_{C},\epsilon_{C}  = \pm1\nonumber\\
\NH(\vv k) &=& \mathcal{K}\NH^{*}(a_{K}\vv k)\mathcal{K}^{-1},\mathcal{K}\mathcal{K}^{*} = \eta_{K}\mathbb{I}, \ \ \eta_{K},a_{K} = \pm1\nonumber.
\end{eqnarray}
The commutation relations between the symmetries are also defined. For example, $\mathcal{C} = \epsilon_{PC} \mathcal{P}\mathcal{C}\mathcal{P}^{T}$. We can define any such $\epsilon_{\mu\nu}$ in this way. Note, in \cite{2018arXiv181210490Z,kawabata2018symmetry} authors always choose an $a_{K} = a_{C} = -1$ for $\mathcal{C}$ and $\mathcal{K}$, this only alters the classification for dimensions greater than zero. We compute topological invariants of the Hamiltonian given the symmetries and their commutation relations via the doubled Hamiltonian formalism. In particular, we construct the Clifford algebra associated with our symmetry generators and compute the extension under an additional generator corresponding to a mass term \cite{kitaev2009periodic}. From the extension problem, we compute the topological invariant by taking the zeroth homotopy group as is done in Hermitian K-theory \cite{Clas1c,Clas2,Clas3,Clas1a,Clas4,Clas5,Clas1b,Slager2015,Rhim2018,PhysRevB.93.245406,2018arXiv181210490Z,kawabata2018symmetry,kawabata2019topological}. We enumerate the relevant cases:
\begin{enumerate}
    \item $h_{z}= h_{y} = 0$, $h_{x} \neq 0$: We have transposition symmetry under inversion, $\mathcal{C} \equiv \mathcal{I}_{y}$,
    $$\mathcal{I}_{y}\NH^{T}(\vv k)\mathcal{I}_{y} = \NH(-\vv k).$$
    We have chiral symmetry under 
    $$\mathcal{I}_{y}\sigma_{z}\NH(\vv k) \sigma_{z}\mathcal{I}_{y} = -\NH(\vv k).$$ 
    Thus, we have $\epsilon_{C} =\eta_{C}=\epsilon_{PC} = 1$, class 10 of \cite{2018arXiv181210490Z}, and a trivial point gap invariant, as the TRS $(\vv k\rightarrow-\vv k)$ trivializes the chiral symmetry \cite{PhysRevLett.120.146402}.
    
    \item $h_{z}= h_{x} = 0$, $h_{y} \neq 0$: We have conjugation symmetry,
    $$\mathcal{I}_{y}\NH^{*}(\vv k)\mathcal{I}_{y} = \NH(-\vv k),$$
    and, transposition symmetry, 
    $$\sigma_{x}\mathcal{I}_{y} \NH^{T}(\vv k)\mathcal{I}_{y} \sigma_{x} = \NH(\vv k).$$
    We have another transposition symmetry under conjugation by $\sigma_{y}$,
    $$\sigma_{y} \NH^{T}(\vv k) \sigma_{y} = -\NH(\vv k).$$   
    We also have chiral symmetry,
    $$\mathcal{I}_{y}\sigma_{z}\NH(\vv k) \sigma_{z}\mathcal{I}_{y} = -\NH(\vv k).$$
    We compose $\mathcal{K}= \mathcal{I}_{y}$ with $\mathcal{C} = \mathcal{I}_{y}\sigma_{x}$ to generate a $\mathcal{Q} = \sigma_{x}$ which inverts momentum, $a_{Q} = -1$. Now, we classify the system with the new symmetry, $\epsilon_{PQ} = \epsilon_{PC} = -1 = a_{C} = a_{Q}$, $\epsilon_{QC} = \epsilon_{C} = \eta_{C} = 1$. We are in class 28 of \cite{2018arXiv181210490Z}, but with the classification $\pi_{0}(\mathcal{R}_{0+d}^{2}) = \mathbb{Z}_{2}^{2}$ instead of $\pi_{0}(\mathcal{R}_{0-d})$ because $a_{C} = 1$. Here $\pi_{0}(\mathcal{R}_{2})$ is just the zeroth homotopy group of $\mathcal{R}$. 
    
    \item $h_{z}= 0$, $h_{x}, h_{y} \neq 0$: We have transposition symmetry under conjugation by $\mathcal{C} = \sigma_{y}$,
    $$\sigma_{y} \NH^{T}(\vv k) \sigma_{y} = -\NH(\vv k).$$
    Similarly, conjugation by $\sigma_{x}$ and inversion, $\mathcal{I}_{y}$, generates a transposition symmetry, 
    $$\sigma_{x}\mathcal{I}_{y} \NH^{T}(\vv k)\mathcal{I}_{y} \sigma_{x} = \NH(\vv k).$$
    And, we still have the traditional chiral symmetry,
    $$\mathcal{I}_{y}\sigma_{z}\NH(\vv k) \sigma_{z}\mathcal{I}_{y} = -\NH(\vv k).$$
    If $C = \sigma_{y}$, then $\epsilon_{C} = \eta_{C} = \epsilon_{PC} = -1$, and if $C = \mathcal{I}_{y}\sigma_{x}$, then $\epsilon_{C} = \eta_{C} =1$, $\epsilon_{PC} = -1$. These two sets of signs are equivalent and present in \cite{2018arXiv181210490Z}, but with momentum inverted.  In this case the point gap invariant is thus $\pi_{0}(\mathcal{R}_{7+d}^{2}) = \mathbb{Z}_{2}^{2}$ for $d =2$ (Class 12).
    
    \item $h_{z} \neq 0$, $h_{x}, h_{y} = 0$: This is equivalent to $h_{y} \neq 0$, $h_{x} = h_{z} = 0$ for this model. The only difference arises when generating an impurity edge in the system for which $k_{y}$ inversion symmetry breaks. C.f. item 2.
    
    \item $h_{z},h_{x} \neq 0$, $h_{y}= 0$: This is equivalent to item 3, $h_{x},h_{y} \neq 0$ and $h_{z} = 0$ with respect to bulk symmetries. Again, there are differences in the presence of an edge.
    
    \item $h_{z}, h_{y} \neq 0$, $h_{x} = 0$: We break chiral symmetry here, and only preserve transposition symmetry via conjugation by $\sigma_{y}$
    $$\sigma_{y}\NH(\vv k)^{T} \sigma_{y} = -\NH(\vv k),$$
    and conjugation symmetry by conjugation with $\sigma_{z}$
    $$\sigma_{z}\NH(\vv k)^{*} \sigma_{z} = -\NH(-\vv k).$$
    Composing the two and taking $\tilde{\mathcal{Q}} = \sigma_{x}$ and rotating $\NH\rightarrow i\NH$, we have $\eta_{Q} = 1$. Then, setting $\mathcal{C} = \sigma_{y}$, we have $\epsilon_{C} = \eta_{C} = \epsilon_{QC} = -1$, and $a_{C} = 1$. The point gap invariant is computed for $d =2$, $\pi_{0}(\mathcal{R}_{2}) = \mathbb{Z}_{2}$.
    
    \item $h_{z},h_{x}, h_{y} \neq 0$: We break chiral symmetry, and only preserve transposition symmetry via conjugation by $\sigma_{y}$
    $$\sigma_{y}\NH(\vv k)^{T} \sigma_{y} = -\NH(\vv k).$$
    Although the symmetry has $a_{C} = 1$ instead of inverting momentum, we simply go to Class 9 of Table II in \cite{2018arXiv181210490Z} and find the point gap invariant to be generated by $\pi_{0}(\mathcal{R}_{7+d})$ instead of $\pi_{0}(\mathcal{R}_{7-d})$, giving $\mathbb{Z}_{2}$ for $d= 2$.
\end{enumerate}
\indent The Bernard Le Claire classification of non-Hermitian non-interacting SPTs is exhaustive \cite{kawabata2019topological,2018arXiv181210490Z,kawabata2018symmetry} in the absence of lattice symmetries. Hence, the doubled Green's function generically encodes all of the topological information in the model. In particular, considering the zeros of doubled Green's function, as mentioned above, guarantees a consistent classification of edge modes and by extension the bulk invariant. With this in mind, we turn to a simple two band model. The total Chern number of all bands in such system is simply zero, $c1 +c2 = 0$. Generalizing this to the doubled Hamiltonian, $c1+c2+c3+c4 = 0$ and $ci -cj = \sum_m z_{m}$,  where $z_{m}$ are the orders of the zeros between bands $i$ and $j$. These fully determine the allowed topological invariant.\\
\subsection{Phase Diagram}\label{chernphaseappend}

\indent Before computing everything from the doubled Green's function which is analytically complicated, we compute the simple case when the single Green's function is well defined. In particular, note that in the phase diagram, panel b of Fig.~\ref{Chern}, we observe a transition at $m_{*}\pm h_{y}$, $m_{*} = 2$. Fortunately, the integral expressions in Supplemental Material \ref{chernpolesappend} are simple for $h_{x} = h_{z} = 0$. We check when only one pole is inside the unit circle of the re-parameterization in \ref{chernpolesappend}, $f(\omega) \pm \gamma(\omega)\in \mathbb{D}$, the unit disk. For only $h_{y} \neq 0$,
\begin{eqnarray}
\Omega^{2} &=&  2+(m^2-h_{y}^{2})-2m\cos(k_y).\nonumber
\end{eqnarray}
For simplicity take $k_{y} = 0$, then 
\begin{eqnarray}
f(0) &=& \frac{-1-(1-m-h_{y})(1-m+h_{y})}{2(1-m-h_{y})}\nonumber
\end{eqnarray}
 and 
\begin{eqnarray}
\gamma &=& \frac{1-(m^{2}-h_{y}^{2}-2m+1)}{2(1-m-h_{y})}\nonumber\\
t_{1,2} &=& \frac{(m-1-h_{y})}{(m^{2}-h_{y}^{2}-2m+1)},\frac{(m^{2}-h_{y}^{2}-2m+1)}{(m-1+h_{y})}.\nonumber
\end{eqnarray}
%\end{widetext}\\
\indent So, $t_{1,2} =\frac{1}{m-1+h_{y}},m-1-h_{y}$. Taking $h_{y} = 0$ returns the same m =2 transition where the pole inside moves outside the unit disk. We thus modify $m_{*}$ into $m_{\pm}=2\pm h_{y}$. This is completely consistent with the doubled Green's function, and with \cite{PhysRevLett.121.136802}.  Note, this does not violate our table. In this case, the single Green's functions are singular everywhere in the non-existent gap (for all momenta), and the doubled Green's function no longer factors as in Eq. \eqref{baredGF}.\\
\indent Now, returning to the doubled Green's function, we compute the transitions numerically, and check the classification in \ref{chernsymappend} above. We see complete agreement with the symmetry classification, see Fig.~\ref{Chern}.

\section{Supplemental Material: Non-Hermtian SSH Model}\label{sshappend}
\begin{figure}
    \centering
    \figuretitle{Non-Hermitian SSH Model Boundary Modes}
    \includegraphics{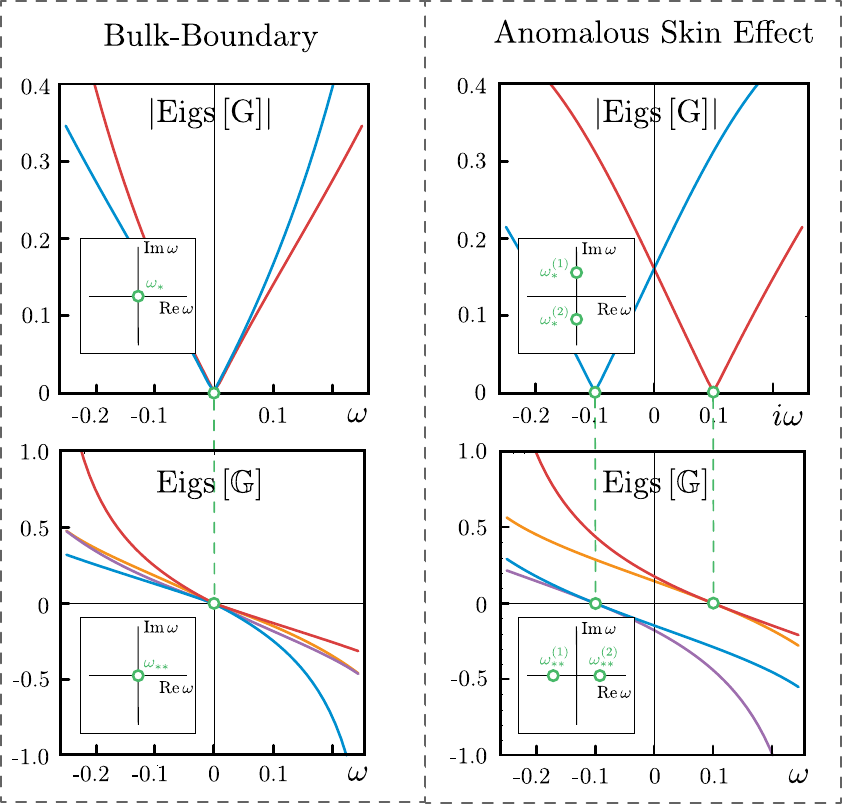}
    \caption{The non-Hermitian SSH model exhibits both types of topological boundary modes. Case I (Left Column), full bulk-boundary correspondence is established (e.g. $m=0.6$, $h_x=h_y=0.1$, $h_z=0$). Case II (Right Column), anomalous skin effect emerges (e.g. $m=0.6$, $h_x=h_y=0$, $h_z=0.1$).}
    \label{SSHfig}
\end{figure}

\indent We consider our arguments in one dimension with the well-studied non-Hermitian version of the {\it Su-Schriefffer-Heeger (SSH) model} \cite{RevModPhys.60.781}.
Specifically, we parameterize the Hermitian SSH model as
\begin{align}
\label{SSH4}
\mathcal{H} =
\vv \xi  \, \vv \sigma 
+ \vv \eta \, \vv \sigma,
\end{align}
\no with 
\begin{align}
\vv \xi  = (\cos k, -\sin k, 0), 
\ \ \ 
\vv \eta  = (m, 0 , 0).
\end{align}
We consider a non-Hermitian version of this Hamiltonian 
\begin{align}
\mathcal H = \vv \xi  \, \vv \sigma 
+ \vv \eta \, \vv \sigma 
+ i \, \vec h \,  \vv \sigma,
\end{align}
\no with  $ \vec h  = (h_x, h_y, h_z) $. Crucially, the model exhibits both chiral and transposition symmetries (even pseudo-Hermiticity in bulk for $h_{x}=h_{y} =0$)\cite{lieu2018topological}. Transposition symmetry implies the existence of an unitary operator, $\sigma_{x}$, relating $\NH$ and $\NH^{T}$, linking right and left Eigenvectors of $\NH$ \cite{Kawabata2018,2018arXiv181210490Z,kawabata2019topological,kawabata2018symmetry}. And, chiral symmetry implies the existence of an unitary operator, $\sigma_{z}$, relating $\NH$ to $-\NH$, linking positive to negative Eigenvalues. In the presence of an edge, inversion symmetry is broken, but the spectrum is still symmetric about both the real and imaginary axes. For each energy, $E$, its counterparts $-E,E^{*},-E^{*}$ are present. And, since there are only two edge modes, $E^{*} = E$ are identified in the edge spectrum, restricting it to the real axis. Therefore, taking $h_{z}=0$ guarantees real edge modes. Taking $h_{z} \neq 0$, while keeping $h_{y}=h_{x} = 0$ preserves the bulk invariant \cite{kawabata2018symmetry}, but pushes the edge modes into the complex plane. Hence, in the topological regime, this simple model exhibits both case I and case II, see Fig.~\ref{SSHfig}.\\
We elaborate some computational details. In particular, we obtain the restricted single Green's function analytically by the same technique used in \cite{Slager2015}.\\
We begin by explicitly computing $\mathcal H ^2$
\begin{align}
\mathcal H ^2 & = (\vv \xi  \, \vv \sigma 
+ \vv \eta \, \vv \sigma 
+ i \, \vec h \,  \vv \sigma)^2  
\nonumber
 \\
 & =  
 \Omega^2  + (\alpha + i \beta ) t + (\alpha^* + i \beta^* ) \frac{1}{t}, 
\end{align}
\no where again $ \Omega^2  =  m^2 +1 - h^2 + 2 i m h_x$ and $\alpha  = m , \ \ \ \beta  =  h_x + i h_y,$.\\
We now proceed to the Fourier transform and put $\vec r_{\bot} = 0$,
\begin{align}
\G (\omega,  \vec r = 0)   
= 
- \frac{1}{(\alpha + i \beta )}
\oint \frac{d t}{2 \pi i }  \frac{\omega + \mathcal H}
{  t^2  - 2 f(\omega) t +c} .
\nonumber
\end{align}
\no with
\begin{align}
f(\omega)  = \frac{\omega^2 - \Omega^2 }
{2 (\alpha + i \beta )},
\ \ \  \ \ \ 
c  = \frac{\alpha^* + i \beta^* }{\alpha + i \beta }.
\end{align}
Thus, we have integral of the same form as in the previous case. We can immediately write
\begin{align}
\label{SSH_GF}
 \G' (\omega)  
  = 
\frac{s (\omega) }{2 \, \gamma (\omega) } 
\begin{pmatrix}
\omega & \alpha^* +  i \beta^{*} + T_1 (\omega)\\
\alpha +  i \beta +\frac{1}{T_2 (\omega)} & \omega
\end{pmatrix} .
\end{align}
\no with $T_{1,2} (\omega) = f (\omega) \pm s(\omega) \gamma (\omega)$ where $s(\omega) = \text{sign} \left( 1 - |t_{1} (\omega) | \right)$ and $t_{1} (\omega) = f (\omega) + \gamma (\omega)$.  
Thus we need only take the determinant of this function to have zeros of the single Green's function and by extension the in-gap bound states.

\end{document}